# Comprehensive characterization of nonlinear viscoelastic properties of arterial tissues using guided-wave optical coherence elastography

Yuxuan Jiang [1, †], Guo-Yang Li [1, 5, †], Ruizhi Wang [3, †], Xu Feng [1, 6, †],

Yanhang Zhang [3, 4, *], Seok-Hyun Yun [1, 2, *]

[1] Harvard Medical School and Wellman Center for Photomedicine, Massachusetts General Hospital, Boston, MA 02114, USA

[2] Harvard-MIT Health Sciences and Technology, Cambridge, MA 02139, USA

[3] Department of Mechanical Engineering, Boston University, Boston, MA 02215, USA

[4] Department of Biomedical Engineering, Boston University, Boston, MA 02215, USA

[5] Currently with Department of Mechanics and Engineering Science, College of Engineering, Peking University, Beijing 100871, China

[6] Currently with Department of Bioengineering, University of Texas at Dallas, TX 75080, USA

[†] These authors contributed equally.

* Correspondence should be addressed to: yanhang@bu.edu (Y.Z.) or syun@hms.harvard.edu (S.H.Y.).

**ABSTRACT**

The mechanical properties of arterial walls are critical for maintaining vascular function under pulsatile pressure and are closely linked to the development of cardiovascular diseases. Despite advances in imaging and elastography, comprehensive characterization of the complex mechanical behavior of arterial tissues remains challenging. Here, we present a broadband guided-wave optical coherence elastography (OCE) technique, grounded in viscoelasto-acoustic theory, for quantifying the nonlinear viscoelastic, anisotropic, and layer-specific properties of arterial walls with high spatial and temporal resolution. Our results reveal a strong stretch dependence of arterial viscoelasticity, with increasing prestress leading to a reduction in tissue viscosity. Under mechanical loading, the adventitia becomes significantly stiffer than the media, attributable to engagement of collagen fibers. Chemical degradation of collagen fibers highlighted their role in nonlinear viscoelasticity. This study demonstrates the potential of OCE as a powerful tool for detailed profiling of vascular biomechanics, with applications in basic research and future clinical diagnosis.

**INTRODUCTION**

The mechanical properties of arterial walls are fundamental to cardiovascular function. Alterations in these properties are associated with a range of vascular pathologies, including hypertension[1], coronary artery diseases[2], and aneurysm[3]. Arterial stiffening[4] and weakening directly impact hemodynamics and can lead to rupture or bulging. Shear stress on arterial walls has been implicated in tortuosity[5,6], buckling[7], dissection[8-10], vasa vasorum circulation[11], and atherosclerotic plaque development[12,13]. Furthermore, changes in tissue nonlinearity and anisotropy reflect underlying structural and compositional remodeling. A non-destructive method capable of characterizing these sophisticated mechanical properties is therefore highly desirable for disease diagnosis and monitoring.

Arterial mechanics primarily derive from the fiber-reinforced structures of the media and adventitia. The adventitia contains a dense network of helically arranged, wavy collagen fibers, which confer tensile strength[14]. The media consists of concentric elastic lamellae composed of elastic fibers[14], interspersed with transmural elastic fibers, collagen, proteoglycans, and smooth muscle cells[15,16]. Elastin and collagen govern tensile responses in low- and high-strain regimes[17,18], respectively, while their orientation determines in-plane anisotropy[19,20]. Non-fibrous components also contribute to shear resistance[9]. The heterogeneous organization of these structures underlies the tissue's anisotropic mechanical response. Age-related stiffening is more prominent longitudinally[21,22], whereas aneurysm exhibit greater circumferential stiffening[23].

Conventional techniques to measure arterial mechanics include planar[21,22] and uniaxial tension tests[24], inflation-extension tests[25,26], shear[9,27], rotated-axes biaxial tests[28], and torsion[11,29]. However, these bulky mechanical techniques are not amenable to *in vivo* application. Ultrasound can monitor arterial diameter and pressure to estimate circumferential modulus[30,31], but suffers from reduced accuracy in small vessels[32]. Pulse wave velocity (PWV), the gold clinical gold standard for stiffness assessment[33], fails to account for mechanical and geometric heterogeneities[34]. Other indices based on pressure waveforms, such as augmentation index[35], central pulse pressure[36], back wave amplitude[37] and harmonic distortion[38], are often compounded by high heart rate, aging, pathology, and pharmacological interventions[32,39,40]. These methods cannot directly or quantitatively assess the intrinsic mechanical properties of arterial walls.

Elastography based on ultrasound[41-43] and magnetic resonance imaging (MRI)[44] enables non-invasive stiffness mapping, but with limited resolution. Optical coherence elastography (OCE), an extension of optical coherence tomography (OCT), offers superior spatial resolution and sensitivity. OCE has been used to measure shear modulus in tissues such as skin[45,46], cornea[47,48], sclera[49], and artery[50]. However, previous OCE methods were generally restricted to shear modulus estimation. Recently, we developed a multi-wave OCE method capable of simultaneously quantifying both tensile and shear moduli in the cornea[51].

In the present study, we further advanced this OCE method and applied it to characterize the mechanical properties of porcine aortas *ex vivo*. Wave propagation velocities were measured in both circumferential and longitudinal directions over a 1-20 kHz frequency range. By analyzing dispersion under biaxial stretch, we extracted nonlinear and anisotropic shear and tensile modulus parameters. To capture viscoelastic behavior, we developed a viscoelastic two-layer guided wave model grounded in our newly proposed generalized acousto-viscoelastic theory[52], enabling quantification of layer-specific mechanical properties in the media and adventitia, including their stretch- and frequency-dependent viscous parameters. Finally, we used selective chemical treatments to remove collagen and investigated the distinct contributions of collagen and elastin to viscoelastic tensile and shear properties. This study demonstrates the utility of OCE for comprehensive mechanical characterization of arterial tissues, with relevance to both basic research and potential clinical applications.

**RESULTS**

*OCE detection of A0- and S0 wave modes*

Porcine aorta samples were cut-open and mounted on a biaxial stretching device (Fig. 1a). Waves were excited using a PZT probe along either the axial or circumferential direction. During measurements, the intimal-media surface faced upward, while the adventitia remained in contact with PBS to prevent dehydration (Fig. 1b). A representative OCT cross-section of the arterial wall is shown in Fig. 1c, where the media and adventitia are distinguishable by their reflectivity and thickness. Based on OCT images, the average wall thickness was $1.53 \pm 0.11$ mm (N = 10), with a media-to-adventitia thickness ratio of approximately 1:1.4. Wave displacements were recorded at excitation frequencies ranging from 1 to 20 kHz. Representative displacement maps at 8 kHz and 16 kHz are shown in Fig. 1d, exhibiting sinusoidal oscillations with exponential decay (Fig. 1e). FFT analysis of the displacement profiles revealed spatial frequency components corresponding to the A0 and S0 modes (Fig. 1f).

Figure 1g illustrates frequency-dependent velocities. At frequencies below 5 kHz, only the quasi-antisymmetric (A0) mode was detectable. Between 5 and 10 kHz, both the A0 and quasi-symmetric (S0) modes were observed, while above 10 kHz, the S0 mode dominated. This is because a wave is most efficiently excited when its half-wavelength approximately matches the contact length of the probe tip. According to Lamb wave theory, the low-frequency S0 mode corresponds to dilatational motion associated with tensile deformation, whereas the A0 mode reflects bending motion involving shear deformation (Fig. 1h)[51]. The phase velocity of the A0 mode increases with frequency and asymptotically approaches to the Scholte wave velocity at the tissue-fluid interface. In contrast, the S0 mode velocity decreases toward the Rayleigh wave limit at the air-tissue interface. Finite element simulations (Fig. 1i) confirmed the presence of both A0 and S0 modes, with asymmetric mode profiles due to the differing boundary conditions on each surface.

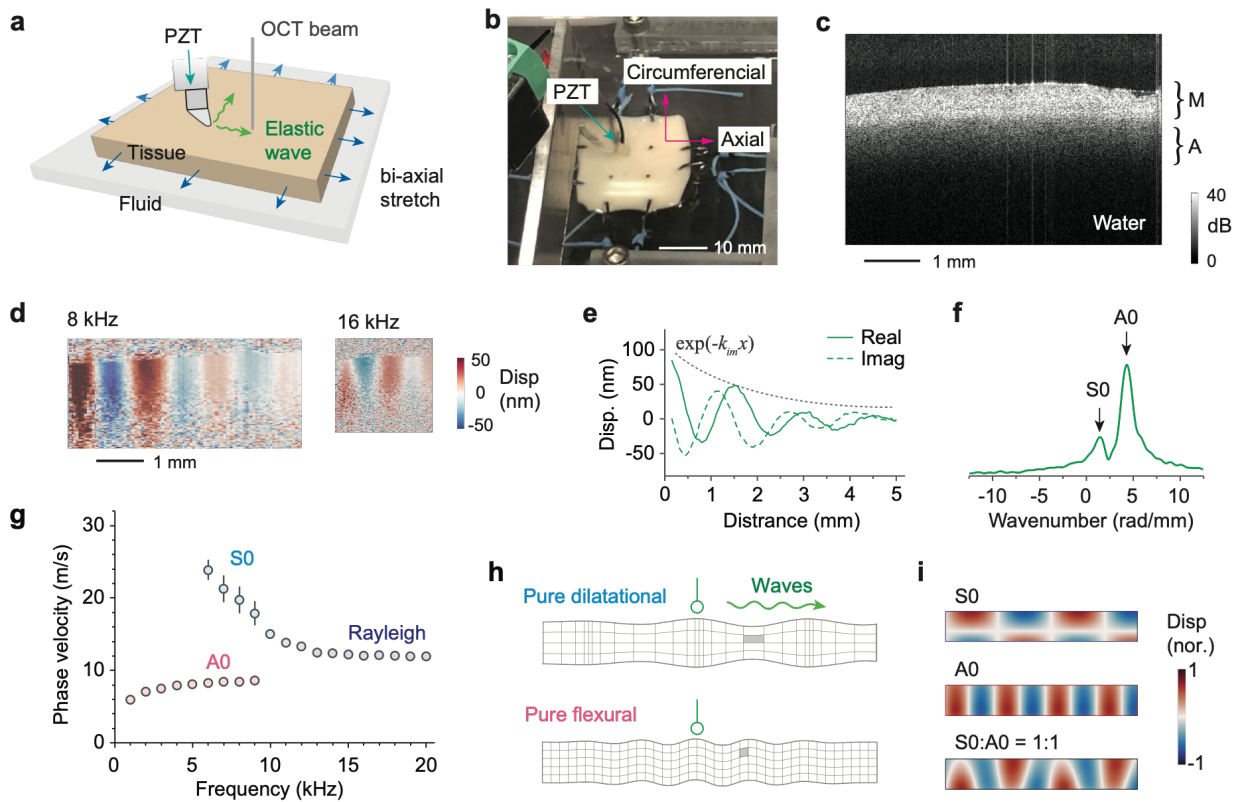

**Fig. 1. S0- and A0-waves in the artery. a**, Schematic of a flattened artery tissue on a water bath to avoid dehydration. The elastic waves are excited by the contact probe and are measured by an OCT beam. **b**, Photograph of the setup. **c**, Typical OCT image of a sample. M: Media. A: Adventitia. **d**, Representative wave motion profile measured in the artery for two different wave frequencies of 8 and 16 kHz. The displacement map (real part) is overlaid on the gray scale optical coherence tomography image. **e**, Displacement extracted along the sample surface at 8 kHz. Solid and dashed lines denote the real and imagery parts of the displacement, respectively. **f**, The displacement is Fourier transformed to wavenumber

space, in which the primary A0 and S0 can be resolved. **g**, Representative experimental data (circles) for phase velocities measured at different frequencies. Two modes are identified between 5 and 10 kHz, corresponding to the A0 and S0 modes. At high frequencies above 10 kHz, only a single mode is reliably detected, which is interpreted as the S0 mode in the limit of Rayleigh surface wave regime. **h**, Schematics of a pure dilatational wave profile (top) and a pure flexural wave displacement (bottom). **i**, Modal shapes of A0 and S0 showing the deformations introduced by the A0 and S0 Lamb waves in the low frequency regime are primarily shear and tensile deformations, respectively. Finite element model simulation results for the modal shapes of the A0 wave, S0 wave, and a combination of the two modes with equal amplitudes.

### *Elastic wave analysis of biaxially stretched tissues*

Wave velocity profiles were measured in both axial and circumferential directions across varying stretch ratios ($\lambda$ = 1.0 to 1.4). As shown in Fig. 2, phase velocities for both A0 and S0 modes increased with stretching. At stretch ratios above 1.2, circumferential S0 mode velocities became unreliable due to low excitation efficiency.

To extract elastic moduli, we modeled the arterial wall as an incompressible elastic plate bordered by air and water. The incremental stress $\mathbf{\Sigma}$ is related to the displacement $\mathbf{u}$ as[53]:

$$\Sigma_{ij} = \mathcal{A}^0_{ijkl} u_{l,k} - \hat{p}\delta_{ij} + p u_{i,j} \tag{1}$$

where $\mathcal{A}^0_{ijkl}$ is the Eulerian elasticity tensor, $p$ is the Lagrange multiplier enforcing incompressibility, and $\hat{p}$ its incremental term. Using a stream function $\psi$, the wave equation becomes (see details in Supplementary Note 1):

$$\alpha \psi_{,xxxx} + 2\beta \psi_{,xxyy} + \gamma \psi_{,yyyy} = \rho(\psi_{,xxtt} + \psi_{,yytt}) \tag{2}$$

Here, $\alpha = \mathcal{A}^0_{xyxy}$, $2\beta = \mathcal{A}^0_{xxxx} + \mathcal{A}^0_{yyyy} - 2\mathcal{A}^0_{xxyy} - 2\mathcal{A}^0_{xyyx}$, $\gamma = \mathcal{A}^0_{yxyx}$. These incremental elastic moduli characterize resistance to shear and in-plane tensile deformation[47]. For $\psi \propto e^{sky} e^{i(kx-\omega t)}$, the characteristic equation becomes:

$$\gamma s^4 - \left(2\beta - \rho \frac{\omega^2}{k^2}\right) s^2 + \alpha - \rho \frac{\omega^2}{k^2} = 0 \tag{3}$$

For a bulk shear wave polarized in the *y*-direction ($s = 0$), this yields $v = \omega/k = \sqrt{\alpha/\rho}$, indicating that $\alpha$ represents the shear modulus. A static plate analysis shows that $2\beta + 2\gamma$ corresponds to the in-plane tensile modulus (see Supplementary Note 6). Applying boundary conditions at the air and water interfaces, we solved the resulting secular equation $\det(\mathbf{M}^e_{5 \times 5}) = 0$ (matrix components in Supplementary Note 1) to fit the dispersion data and extracted $\alpha$, $\beta$, and $\gamma$. The derived moduli are listed in Table 1, and corresponding fit curves are plotted in Fig. 2.

The elastic model captured overall trends but showed discrepancies— particularly for the A0 mode below 5 kHz at $\lambda$ = 1-1.2 and for the S0 mode above 15 kHz. Notably, in unstretched samples ($\lambda$ = 1), the S0 velocity increased with frequency, whereas the elastic model predicted a monotonous decrease toward the Rayleigh surface wave limit. These deviations suggest viscoelastic contributions, addressed in the next section.

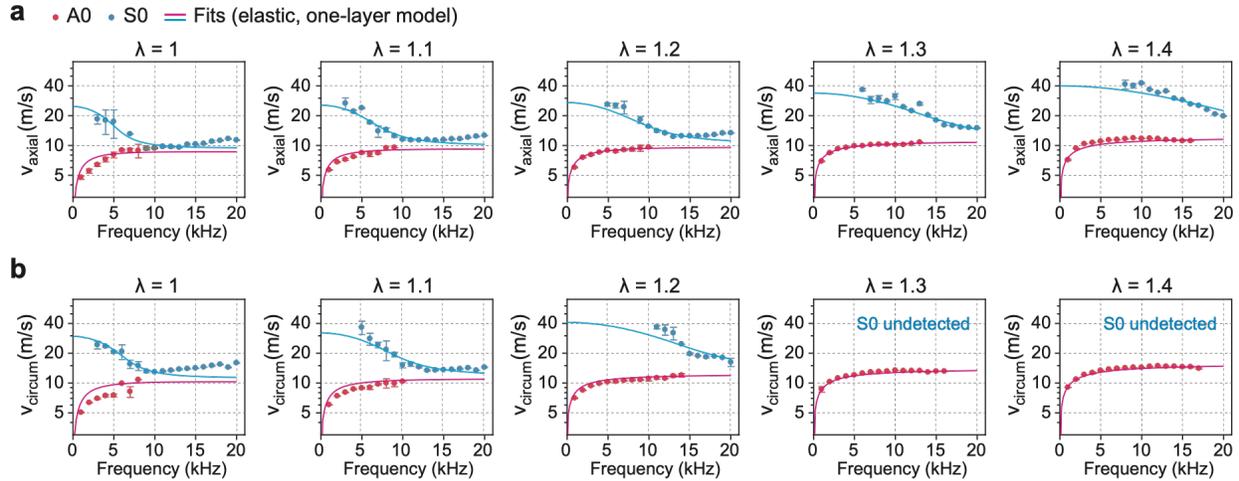

**Fig. 2. Phase velocities in axial and circumferential directions, and their fitting curves using the single-layer elastic model. a**. Axial dispersion relations of A0 and S0 modes measured at varying stretch ratios. **b**. Circumferential dispersion relations of A0 and S0 when stretch ratio $\lambda$ increases from 1 to 1.4. Markers: experiments. Lines: fitting curves using the single-layer elastic model. The parameters are listed in Table 1.

**Table 1.** Measured stretch-dependent elastic moduli from the single-layer elastic model (1-20 kHz)

|  | Modulus (kPa) | $\lambda$ = 1.0 | $\lambda$ = 1.1 | $\lambda$ = 1.2 | $\lambda$ = 1.3 | $\lambda$ = 1.4 |
|---|---|---|---|---|---|---|
| Axial | Shear, $\alpha$ | 92 ± 1.0 | 101 ± 0.2 | 106 ± 0.1 | 131 ± 0.4 | 171 ± 0.3 |
|  | Tensile, $2\beta + 2\gamma$ | 610 ± 45 | 650 ± 190 | 740 ± 120 | 1140 ± 130 | 1600 ± 220 |
| Circum. | Shear, $\alpha$ | 131 ± 0.1 | 141 ± 0.4 | 163 ± 1.3 | 206 ± 0.7 | 260 ± 1 |
|  | Tensile, $2\beta + 2\gamma$ | 876 ± 4 | 1030 ± 30 | 1700 ± 125 | - | - |

### *Viscoelastic single-layer wave model analysis*

To account for frequency-dependent behavior of arterial tissues[54], we incorporated a Kelvin-Voigt fractional derivative (KVFD) viscoelastic model[55,56]. In this formulation, a viscoelastic "spring-pot" element operates in parallel with an elastic spring (Fig. 3a). The spring-pot is defined by a complex, frequency-dependent parameter:

$$\Omega = \eta(i\omega)^\delta \quad (4)$$

where $\delta$ is the fractional order, and $\eta$ (unit: $s^\delta$) denotes the relative strength of the spring-pot viscosity compared to the elasticity of the accompanying spring. When $\delta = 1$, the model reduces to the classical Kelvin-Voigt model, where $\Omega = i\eta\omega$. In the linear regime under negligible pre-stress, the parallel combination of a spring-pot and a purely elastic spring with storage modulus $\mu$ yields a complex dynamic modulus of $(1 + \Omega)\mu$. Note that $\Omega = 0$ in response to static stress (since $\omega = 0$). At equilibrium with static pre-stress, the viscous response of the spring-pot has fully relaxed. However, when additional dynamic strain is introduced by acoustic waves, the spring-pot can contribute significantly to the material response.

Incorporating the KVFD model into the pre-stressed, dynamic-strain regime, the incremental stress tensor $\mathbf{\Sigma}$ is modified to[52]:

$$\Sigma_{ij} = G\mathcal{A}^0_{ijkl}u_{l,k} - \hat{q}\delta_{ij} + qu_{i,j} - G\hat{Q}\delta_{ij} + GQu_{i,j} - \Omega\sigma^e_{Dik}u_{j,k} \quad (5)$$

Here, $G = 1 + \Omega$, and $q$ is the Lagrange multiplier with its increment $\hat{q}$. $Q = \sigma^e_{ii}/3$ and $\hat{Q}$ is its increment. The elastic Cauchy stress-strain relation is $\boldsymbol{\sigma}^e = (\partial W/\partial \mathbf{F})\mathbf{F}^\mathrm{T}$, and the deviatoric elastic stress is $\boldsymbol{\sigma}^e_D = \boldsymbol{\sigma}^e - Q\mathbf{I}$. When $\Omega = 0$, Eq. (5) reduces to the elastic form given by Eq. (1), as $q + Q$ corresponds to the original Lagrange multiplier $p$. Inserting this modified stress expression into the wave equation and applying the stream function $\psi$, we obtain (see details in Supplementary Note 2):

$$(G\gamma - \Omega\sigma^e_{Dyy})s^4 + \left[\rho\frac{\omega^2}{k^2} - 2G\beta + \Omega(\sigma^e_{Dxx} + \sigma^e_{Dyy})\right]s^2 + \left(G\alpha - \Omega\sigma^e_{Dxx} - \rho\frac{\omega^2}{k^2}\right) = 0 \quad (6)$$

Applying the same boundary conditions used in the elastic model, we solved the corresponding secular equation $\det(\mathbf{M}^V_{5\times 5}) = 0$, where the matrix components are detailed in Supplementary Note 2. By fitting this model to the experimentally measured dispersion curves, we extracted both the elastic parameters $\alpha$, $\beta$ and $\gamma$ and the viscoelastic parameters $\eta$ and $\delta$. The results are summarized in Table 2, with fitting curves shown in Fig. 3b-c.

Compared to the purely elastic model, the viscoelastic model provided a substantially improved fit, particularly in capturing the dispersion behavior of the A0 mode at low frequencies. The inclusion of viscosity resulted in a more gradual, yet continuous, increase in A0 phase velocity with frequency. However, some mismatch remained for the S0 mode at high frequencies. This discrepancy could not be resolved solely by increasing viscosity, as doing so introduced errors in other frequency regions. These limitations are further addressed in the next section.

From Eq. (6) for $s = 0$, we find that the bulk shear modulus is equal to $G\alpha - \Omega\sigma_{Dxx}^e$, where $\sigma_{Dxx}^e$ is typically an order of magnitude smaller than $\alpha$ (see Supplementary Note 7). To better understand the role of viscoelasticity, we compared the shear modulus $\alpha$ obtained from the elastic model (Fig. 3d) with the real and imaginary parts of $\alpha G$ obtained from the viscoelastic model (Fig. 3e). The real-part of $\alpha G$ represents the storage modulus, reflecting elastic energy retention, and was in close agreement with the elastic model around 10 kHz. As expected for fiber-reinforced tissues, the storage modulus increased with stretch. The imaginary part of $\alpha G$, corresponding to the loss modulus, quantifies viscous energy dissipation and was found to decrease with increasing stretch ratio. This stretch-dependent reduction in loss modulus indicates that arterial tissues exhibit a transition toward more elastic and less viscous behavior as they are deformed.

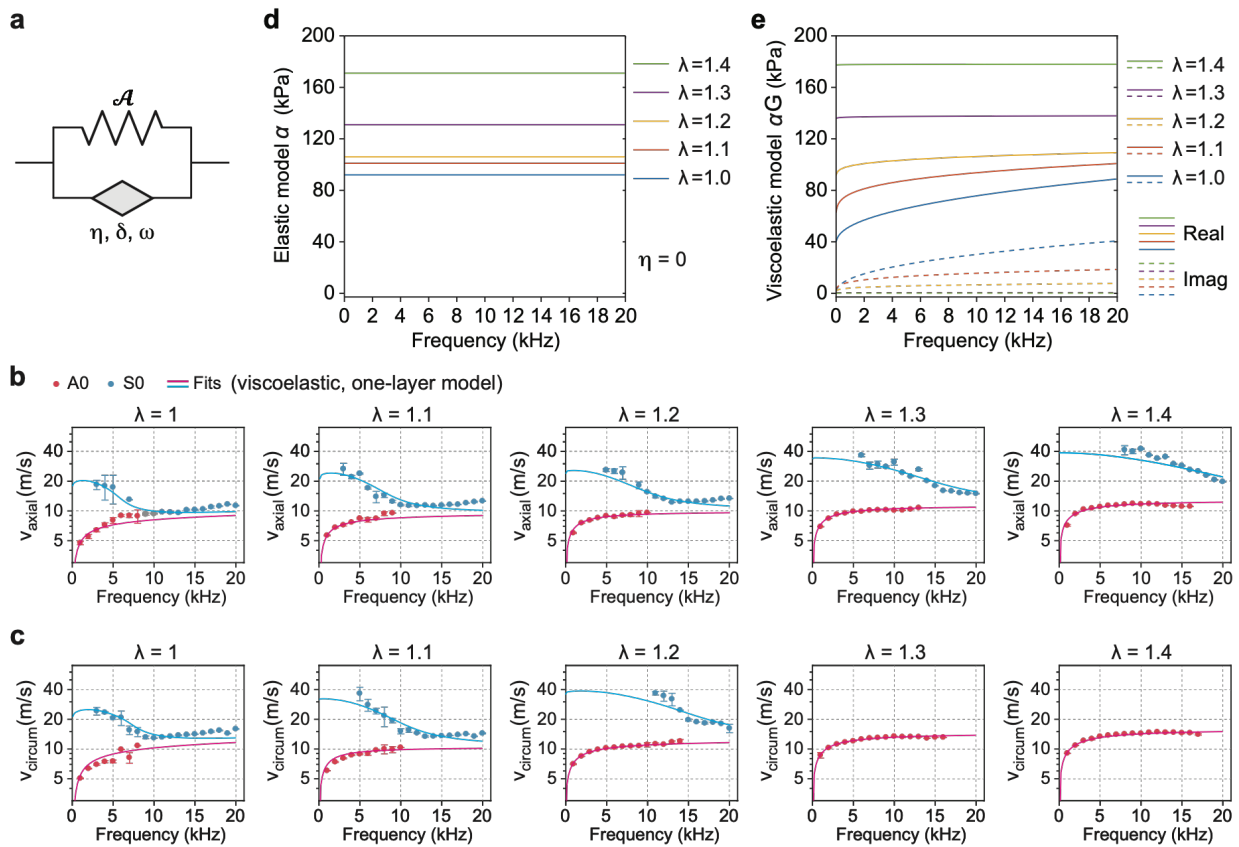

**Fig. 3. Single-layer viscoelastic model analysis of experimental data. a**, Schematic of the KVFD model. **b**, Axial dispersion relations of A0 and S0 modes. **c**, Circumferential dispersion relations of A0 and S0 modes. Markers: experiments. Lines: fitting curves using the single-layer viscoelastic model. The parameters are listed in Table 2. **d**, Axial $\alpha$ parameter values derived with the pure elastic model (from Table 1). **e**, Product of $\alpha$ times $1 + \eta(i\omega)^\delta$ obtained from the axial values in Table 2. Solid curves: real values, Dashed curves: imaginary values.

**Table 2.** Measured modulus parameters from the single-layer viscoelastic model (1-20 kHz)

|  |  | $\lambda = 1.0$ | $\lambda = 1.1$ | $\lambda = 1.2$ | $\lambda = 1.3$ | $\lambda = 1.4$ |
|---|---|---|---|---|---|---|
| Axial | $\alpha$ (kPa) | 38 ± 12 | 56 ± 12 | 87 ± 29 | 132 ± 32 | 176 ± 18 |
|  | $2\beta + 2\gamma$ (kPa) | 320 ± 60 | 420 ± 140 | 570 ± 110 | 1140 ± 170 | 1500 ± 240 |
|  | $\eta$ ($\times 10^{-3} s^\delta$) | 11 ± 4 | 46 ± 14 | 23 ± 14 | 22 ± 14 | 5.5 ± 1.7 |
|  | $\delta$ | 0.43 ± 0.14 | 0.25 ± 0.14 | 0.21 ± 0.13 | 0.06 ± 0.03 | 0.06 ± 0.01 |
| Circum. | $\alpha$ (kPa) | 56 ± 6 | 109 ± 45 | 122 ± 48 | 215 ± 19 | 270 ± 9 |
|  | $2\beta + 2\gamma$ (kPa) | 440 ± 90 | 1000 ± 180 | 1300 ± 290 | - | - |
|  | $\eta$ ($\times 10^{-3} s^\delta$) | 9 ± 5 | 4 ± 2 | 11 ± 6 | 14 ± 5 | 1.2 ± 0.1 |
|  | $\delta$ | 0.46 ± 0.07 | 0.29 ± 0.13 | 0.30 ± 0.11 | 0.08 ± 0.05 | 0.05 ± 0.01 |

### *Viscoelastic two-layer wave model analysis*

The media and adventitia exhibit distinct structural compositions and mechanical properties[14], with adventitia containing more collagen-rich, highly anisotropic fibers, and the media dominated by elastic lamellae. To account for this heterogeneity, we extended the viscoelastic wave model to a two-layer configuration. Each layer was assigned independent elastic moduli, denoted $\alpha_1$, $\beta_1$, and $\gamma_1$ for the media and $\alpha_2$, $\beta_2$, and $\gamma_2$ for the adventitia, while the viscous parameters $\eta$ and $\delta$ were assumed to be identical across layers to reduce the number of free parameters. Continuity conditions for displacement and stress were applied at the media-adventitia interface, in addition to the boundary conditions at the air-tissue and fluid-tissue interfaces. These interfacial conditions yielded a secular equation of the form: $\det(\mathbf{M}_{9\times9}^v) = 0$ with matrix components detailed in Supplementary Note 3.

Figure 4 presents the fitted dispersion curves. The two-layer viscoelastic model successfully captured key trends in the high-frequency behavior of both A0 and S0 modes: specifically, the upward trend of the S0 mode at $\lambda = 1$ and 1.1, and the downward trend of the A0 mode at higher stretches ratios ($\lambda = 1.3$ and 1.4). These behaviors reflect the evolving mode shape with increasing frequency. Depending on the modulus ratio of the two layers, the asymptotic velocities of the S0 and A0 modes differ (see details in Supplementary Note 8). In the stress-free state ($\lambda =$ 1), where the media is slightly stiffer than the adventitia ($\alpha_2/\alpha_1 \approx 0.8$), the S0 mode approaches the shear wave velocity of the adventitia, and the A0 mode approaches the fluid–adventitia interface, resembling a Scholte wave limit. Under stretched conditions ($\lambda \geq 1.1$), the adventitia becomes significantly stiffer than the media ($\alpha_2/\alpha_1 > 1.4$); thus, the S0 mode approaches the shear wave velocity of the media, and the A0 mode tends toward the air–media interface, resembling the Rayleigh wave limit. Importantly, these trends could not be reproduced by a two-layer elastic model with $\eta = 0$, which yielded dispersion curves similar to those of the single-layer elastic case (Supplementary Fig. S1 and Supplementary Note 4). This confirms that the observed high-frequency behaviors result from the combined effects of viscosity and spatially varying stiffness.

Fitted viscoelastic parameters are summarized in Table 3 and plotted in Fig. 5. Both the shear and tensile moduli increased with stretch. A particularly sharp increase in adventitial tensile modulus was observed with $\lambda > 1.1$ (Fig. 5e), in agreement with previous biaxial tensile studies[22]. In the unstressed condition ($\lambda = 1$), the media exhibited greater stiffness than the adventitia (Fig. 5c, f). However, under stretch, the adventitia stiffened more rapidly, eventually surpassing the media in stiffness. Additionally, circumferential elastic moduli were consistently greater than axial moduli under tension.

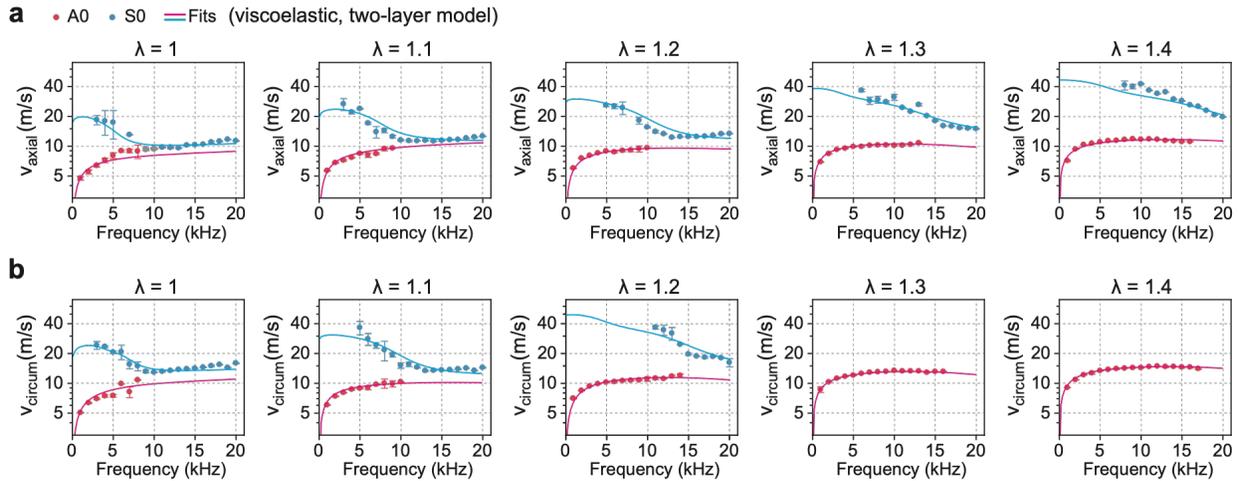

**Fig. 4. Two-layer viscoelastic analysis of axial and circumferential velocities. a**. Axial dispersion relations of A0 and S0 modes. **b**, Circumferential dispersion relations of A0 and S0 modes. Markers: experiments. Lines: fitting curves using the two-layer viscoelastic model.

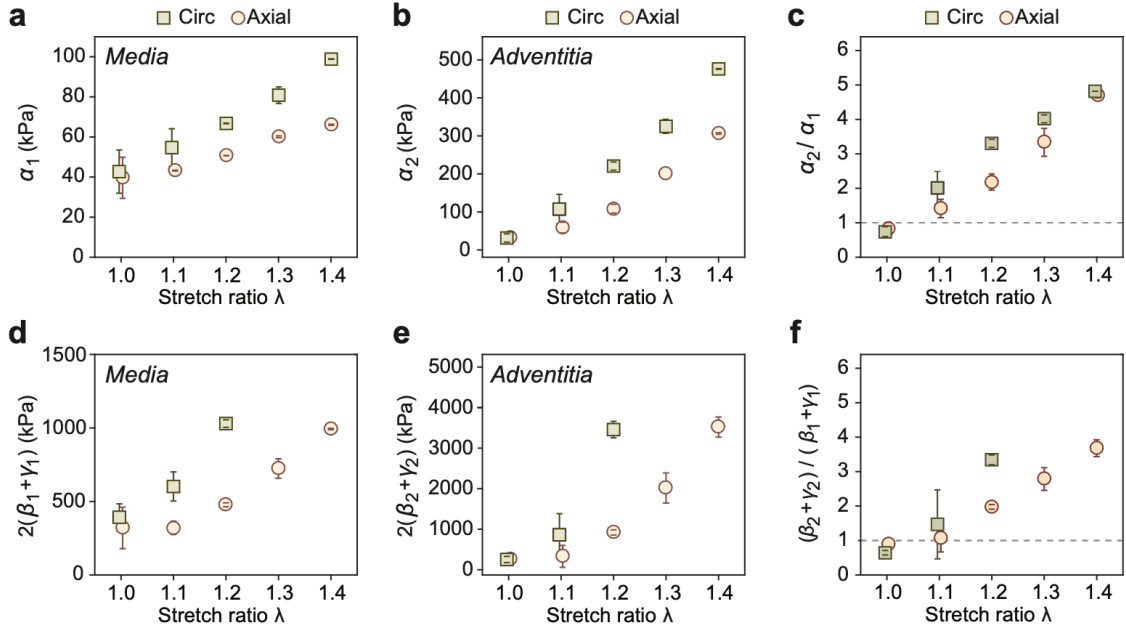

**Fig. 5**. **Shear and tensile moduli of the media and the adventitia**. **a**, Axial and circumferential shear moduli of the media as functions of the stretch ratio $\lambda$. **b**, Axial and circumferential shear moduli of the adventitia. **c**, The ratio of adventitial over medial shear moduli along both axial and circumferential directions. **d**, Axial and circumferential tensile moduli of the media. **e**, Axial and circumferential tensile moduli of the adventitia. **f**, The ratio of adventitial over medial tensile moduli along both axial and circumferential directions.

**Table 3.** Measured modulus parameters from the two-layer viscoelastic model (1-20 kHz)

|  |  | $\lambda = 1.0$ | $\lambda = 1.1$ | $\lambda = 1.2$ | $\lambda = 1.3$ | $\lambda = 1.4$ |
|---|---|---|---|---|---|---|
| Axial, intima-media | $\alpha$ (kPa) | 40 ± 10 | 43 ± 0.2 | 51 ± 0.1 | 60 ± 0.4 | 66 ± 0.3 |
|  | $2\beta + 2\gamma$ (kPa) | 300 ± 140 | 330 ± 40 | 480 ± 10 | 720 ± 70 | 1000 ± 5 |
| Axial, adventitia | $\alpha$ (kPa) | 32 ± 15 | 60 ± 14 | 110 ± 10 | 200 ± 26 | 310 ± 2 |
|  | $2\beta + 2\gamma$ (kPa) | 270 ± 140 | 340 ± 270 | 930 ± 60 | 2020 ± 370 | 3500 ± 250 |
| Axial (media & adventitia) | $\eta$ ($\times 10^{-3} s^{\delta}$) | 10 ± 6 | 25 ± 0.5 | 25 ± 1 | 20 ± 2.6 | 9 ± 3 |
|  | $\delta$ | 0.45 ± 0.06 | 0.36 ± 0.10 | 0.25 ± 0.06 | 0.15 ± 0.03 | 0.09 ± 0.02 |
| Circum., intima-media | $\alpha$ (kPa) | 43 ± 11 | 55 ± 9 | 67 ± 0.2 | 81 ± 4 | 99 ± 0.1 |
|  | $2\beta + 2\gamma$ (kPa) | 390 ± 90 | 600 ± 100 | 1030 ± 26 | - | - |
| Circum., adventitia | $\alpha$ (kPa) | 32 ± 11 | 110 ± 40 | 220 ± 10 | 325 ± 20 | 48 ± 1 |
|  | $2\beta + 2\gamma$ (kPa) | 250 ± 76 | 880 ± 520 | 3450 ± 200 | - | - |
| Circum. (media & adventitia) | $\eta$ ($\times 10^{-3} s^\delta$) | 15 ± 2 | 21 ± 4 | 25 ± 0.2 | 15 ± 0.2 | 1.3 ± 0.1 |
| | $\delta$ | 0.46 ± 0.06 | 0.30 ± 0.05 | 0.17 ± 0.02 | 0.21 ± 0.03 | 0.05 ± 0.03 |

*Stretch-dependent viscosity parameters of arterial tissues*

Figure 6 summarizes the viscoelastic parameters $\eta$ and $\delta$ extracted from the two-layer model and listed in Table 3. The amplitude parameter $\eta$ increases with stretch up to $\lambda$ = 1.2, then decreases at higher stretch ratios. The underlying mechanistic basis for this non-monotonic trend remains unclear but may reflect microstructural changes in fiber alignment and fluid redistribution during deformation. In contrast, the fractional order $\delta$ decreases consistently with increasing stretch, remaining within a range of 0 to 0.5—comparable to prior reports (0.1-0.3) from uniaxial stress relaxation experiments[54].

Representing the dynamic modulus of the spring-pot, both the real and imaginary components of $\alpha\Omega$ and $(2\beta + 2\gamma)\Omega$ were found to decrease with increasing stretch. This trend indicates a reduction in both energy storage and dissipation contributed by the spring-pot component. The loss tangent, defined as the ratio of the loss modulus to the storage modulus, also decreases with stretch. Collectively, these findings demonstrate that arterial viscoelasticity is highly deformation-dependent with tissues exhibiting reduced viscosity under increasing tension.

The attenuation of acoustic waves, visible in the wave profile (Fig. 1f), also reflects this viscoelastic behavior. To quantify attenuation, we fit the displacement amplitude profiles to an exponential decay model of the form, $e^{-k_{\text{im}}x}$, where $k_{\text{im}}$ is the imaginary part of the wavenumber. The measured attenuation coefficients (Fig. 6c) increased with frequency, in agreement with the rising loss modulus shown in Fig. 3e. Notably, the attenuation decreased as the stretch ratio increased, further supporting the observation that arterial tissues become less dissipative and more elastically dominated as they are stretched.

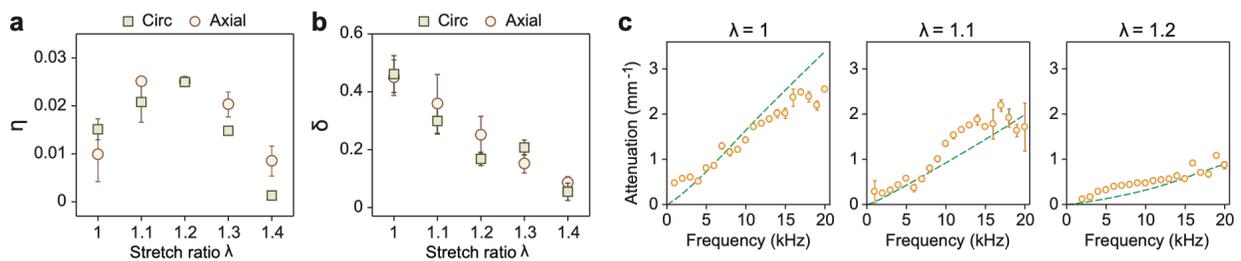

**Fig. 6. Viscoelasticity of the artery. a**, Axial and circumferential viscous parameters $\eta$ with respect to the stretch ratio. **b**, Axial and circumferential viscous parameters $\delta$ (fractional order) with respect to the stretch ratio. **c**, Wave attenuation in the axial direction, with $\lambda$ varying from 1 to 1.2. Markers: experiments. Dashed lines: two-layer model-predicted attenuation of the A0 mode using previously obtained viscoelastic parameters. The attenuation curves for the S0 mode are similar (Supplementary Fig. S2).

*Effects of removal of collagen fibrils*

To investigate the roles of collagen and elastic fibers we treated arterial tissues with cyanogen bromide (CNBr), a process that degrades and removes collagen fibers, cellular components, and other extracellular matrix elements, while largely preserving the elastin fiber network. This treatment reduced the wall thickness from $1.53 \pm 0.11$ mm to $1.19 \pm 0.19$ (N = 5). Figure 7a shows representative circumferential velocity measurements before and after CNBr treatment. Both shear and tensile moduli were substantially reduced following treatment.

Due to the loss of collagen, the samples could be stretched up to a stretch ratio of $\lambda = 1.1$, beyond which mechanical failure occurred at the hooks. From the measured velocity data with the one-layer viscoelastic model (Supplementary Fig. S3), we derived the corresponding mechanical parameters. Table 4 and Figure 7b-e summarizes these results. The shear and tensile moduli of the treated tissues, now dominated by the elastin network, exhibited a substantial decrease compared to those of the intact sample. Circumferential moduli remained slightly higher than axial values, which reflects the anisotropy of the elastin network[17,20]. The viscous parameters were generally comparable to those of the intact arteries. The frequency-dependent axial shear and tensile moduli, $\alpha G$ and $2(\beta + \gamma)G$ are plotted in Fig. 7 f-g. Compared to intact arteries, the treated tissues exhibited considerable stiffening in storage modulus at relatively low strain levels (5-10%). In contrast, changes in loss moduli were modest, if not negligible.

We applied the GOH constitutive model (see Eq. (10)), which is widely used to describe fiber-reinforced cardiovascular tissues. The model parameters, estimated from the measured axial and circumferential elastic moduli (Tables 2-4), are listed in Table 5. In intact samples, the parameter $k_1$ in the adventitia was higher than that in the media, consistent with the higher collagen content in the adventitia. Following collagen removal, both $\mu_0$ and $k_1$ decreased, and notably, the nonlinear exponent $k_2$ was reduced by more than half.

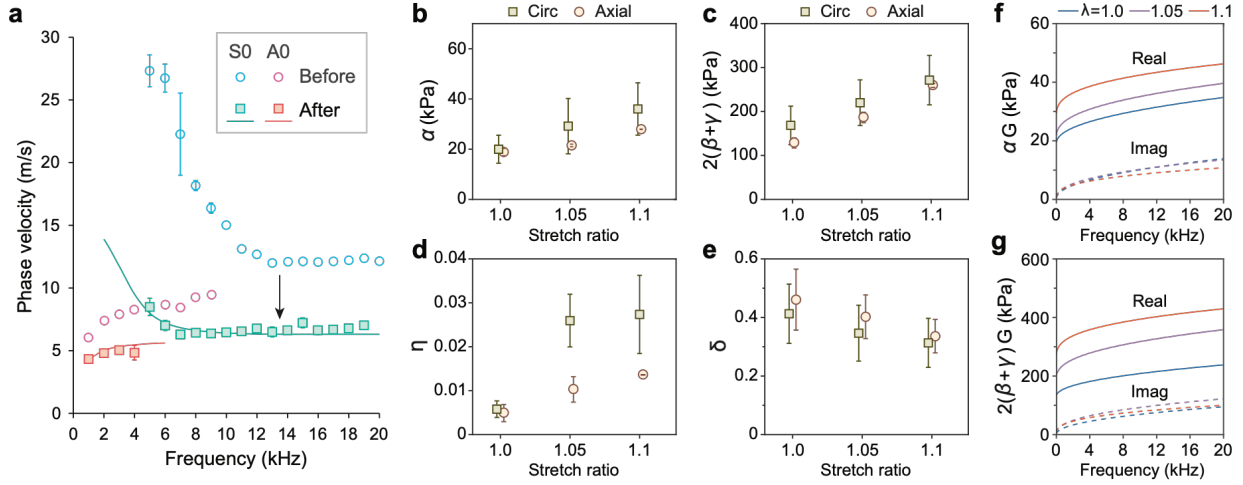

**Fig. 7. Viscoelastic properties of arterial tissues after treating with CNBr. a**. Representative dispersion relations for an arterial tissue before and after treatment. A 10% strain ($\lambda$ = 1.1) was applied in both cases. Curves: fitting with the viscoelastic one-layer model. **b-e**, Viscoelastic parameters after treatment at different stretch ratios: Axial and circumferential shear moduli (b), tensile moduli (c), with respect to the stretch ratio, viscous parameters $\eta$ (d), and fractional orders δ (e). **f**, Product of $\alpha$ times $G = 1 + \eta(i\omega)^\delta$ obtained from the axial values. **g**, Product of axial $2(\beta + \gamma)$ times $G$. Solid curves: real values, Dashed curves: imaginary values.

**Table 4**. Viscoelastic parameters measured on collagen-degraded tissues (1-20 kHz).

|  |  | $\lambda$ = 1.0 | $\lambda$ = 1.05 | $\lambda$ = 1.1 |
|---|---|---|---|---|
| Axial | Shear, $\alpha$ (kPa) | 19 ± 1 | 21 ± 0.5 | 28 ± 0.1 |
|  | Tensile, $2\beta + 2\gamma$ (kPa) | 130 ± 12 | 190 ± 12 | 260 ± 5 |
|  | $\eta$ ($\times 10^{-3} s^\delta$) | 5 ± 2 | 10 ± 3 | 14 ± 0.1 |
|  | δ | 0.46 ± 0.10 | 0.40 ± 0.07 | 0.34 ± 0.06 |
| Circum. | Shear, $\alpha$ (kPa) | 20 ± 6 | 29 ± 11 | 36 ± 10 |
|  | Tensile, $2\beta + 2\gamma$ (kPa) | 170 ± 44 | 220 ± 52 | 270 ± 56 |
|  | $\eta$ ($\times 10^{-3} s^\delta$) | 6 ± 2 | 26 ± 6 | 27 ± 9 |
|  | δ | 0.41 ± 0.10 | 0.35 ± 0.10 | 0.31 ± 0.08 |

**Table 5.** Derived constitutive parameters.

| GOH constitutive parameters | $\mu_0$ (kPa) | $k_1$ (kPa) | $k_2$ | $\varphi$ (°) | $\kappa$ |
|---|---|---|---|---|---|
| Single-layer model (from Table 2) | 32 ± 0.3 | 110 ± 0.2 | 4.2 ± 0.3 | 20 ± 0.2 | 0.19 ± 0.1 |
| Media (from Table 3) | 36 ± 0.1 | 103 ± 0.3 | 3.2 ± 0.1 | 35 ± 0.1 | 0.15 ± 0.1 |
| Adventitia (from Table 3) | 36 ± 0.1 | 120 ± 0.1 | 10.1 ± 0.1 | 30 ± 0.1 | 0.16 ± 0.1 |
| After CNBr treatment (from Table 4) | 20 ± 0.1 | 71 ± 0.1 | 2.0 ± 0.1 | 38 ± 0.1 | 0.15 ± 0.1 |

**DISCUSSION**

We have presented a comprehensive characterization of the mechanical properties of arterial tissues using a broadband OCE technique. The high spatial resolution (~10 µm) and vibration sensitivity (~1 nm per A-line) of OCE enabled precise detection of mechanical wave propagation in the tissue. The wave velocities, measured over a broad frequency (1-20 kHz), exhibited rich spectral features, allowing us to extract various mechanical parameters, including shear and tensile elastic moduli as well as viscous coefficients. This acousto-elastic analysis is grounded in continuum mechanics theory and augmented by our novel viscoelastic model framework. Our analytic approach leveraged the layered, two-dimensional architecture of the vascular wall. The tissue supports two distinct types of guided acoustic waves: quasi-antisymmetric (A0) and quasi-symmetric (S0) modes. The OCE system was optimized for efficient excitation and detection of both modes. At frequencies below 5 kHz, the A0 mode was predominant, with dispersion profile providing information about the viscoelastic shear modulus. At higher frequencies (>10 kHz), the S0 mode dominated. The extrapolated dispersion of the S0 mode to lower frequencies yielded estimates of the tensile modulus, while its asymptotic behavior above 15 kHz enabled the estimation of layer-specific viscoelastic parameters.

Our measurements revealed key features of arterial wall mechanics, including anisotropy, nonlinearity, viscoelasticity, and layer-inhomogeneity across physiologically relevant stretch levels in the 1-20 kHz frequency range. Both shear and tensile moduli increased with stretch, with circumferential values consistently exceeding axial ones. The adventitia exhibited greater stiffness than the media under prestressed conditions, highlighting the load-bearing role of collagen fibers [22]. We compared our experimentally derived moduli with literature values obtained from conventional mechanical testing methods[57,58] (see Supplementary Note 9). Overall, the trends are consistent: circumferential moduli are higher than axial moduli, and the adventitia-to-media modulus ratio increases markedly under stretch—from below unity in the unloaded state to values significantly greater than one under physiological tension. In older individuals, aortic stiffness has been reported to be greater in the longitudinal direction than in the circumferential direction[21]. The current OCE technique holds potential for investigating age-related changes in the anisotropy of human arteries and merits further exploration.

An important finding of our study is the stretch-dependent modulation of arterial viscoelasticity. With increasing prestress, we observed a consistent decrease in both wave attenuation and the fractional orders of the viscoelastic model, indicating a shift toward more elastic behavior. Notably, the imaginary components of $\alpha G$ and $(2\beta + 2\gamma)G$ —corresponding to the shear and tensile loss moduli, respectively, and thus indicative of viscous energy dissipation— decreased with increasing stretch ratio. This reduction in loss modulus suggests that arterial tissues transition toward a more elastic, energy-efficient state under physiological loading, potentially optimizing function during cyclic deformation. While nonlinear viscoelasticity in arteries and other biological tissues has been investigated previously[59-63], our quantitative findings offer new insight into this behavior. These results have important implications for constitutive modeling of arteries and may inform the design of bioinspired materials or therapeutic interventions for vascular disease. Further studies are warranted to elucidate the underlying biophysical mechanisms and to determine whether similar viscoelastic trends are observed in other tissue types.

Following collagen removal by CNBr treatment, the treated samples exhibited substantially reduced elastic moduli ($\alpha$, $\beta$, and $\gamma$), which is in agreement with prior studies using enzymatic digestion[64,65]. These results are consistent with previous findings that elastin and collagen fibers predominantly govern tensile behavior in the low- and high-strain regimes, respectively[17,18]. Interestingly, despite the loss of collagen, the complex viscoelastic parameter $\Omega$ (the 'spring-pot') remained comparable to that of intact tissues. This suggests that collagen fibers contribute minimally to viscoelastic damping at low strains, but may play a nonlinear role in viscous dissipation under larger deformations.

The demonstrated OCE technique and acousto-viscoelastic model have several limitations. First, it characterizes mechanical properties at frequencies ranging from 1 to 20 kHz, which are substantially higher than the physiologically relevant frequencies near 1 Hz. While the KVFD model allows extrapolation of viscoelastic moduli to lower frequencies, the accuracy of this extrapolation remains to be validated. Second, the acoustic wavelengths in our measurements ranged from approximately 2 mm (A0 mode) to 7 mm (S0 mode). Since these wavelengths are shorter than the radius of curvature of aortas, our measurements on flattened samples reasonably approximate those in intact cylindrical vessels. However, for lower-frequency waves or smaller-diameter vessels, curvature effects my significantly alter wave propagation, which should be considered in modeling[66]. Third, in our experimental setup, the intima surface was exposed to air to facilitate wave excitation, whereas physiologically it is in contact with blood, and the adventitia is surrounded by soft connective tissues. These in vivo boundary conditions differ from our experimental configuration and are expected to affect the observed mechanical responses, even if the intrinsic properties of the medial and adventitial layers remain the same. Fourth, the accuracy of velocity measurements is constrained by the efficiency of wave excitation and the optical signal-to-noise ratio. Incomplete dispersion curves and parameter interdependence in the fitting model contribute to uncertainties in some mechanical parameters, which exceeded 50%. Future system optimization may help reduce these errors.

Lastly, a major limitation of the current technique is the need for a contact probe to excite guided waves. To overcome this, we aim to implement focused ultrasound, similar to that used in shear-wave ultrasound elastography[67]. Ultrasound allows adjustable focal size, potentially enabling more uniform excitation across frequency modes. Importantly, a non-contact ultrasound-based system would enhance the translation of OCE for clinical applications. Both ultrasound and optical beams could be delivered through intravascular fiber-optic catheters[68-71], enabling simultaneous structural and mechanical assessment of arterial walls. In addition, handheld devices could be developed for noninvasive access to carotid arteries in the head and neck[72], further broadening the utility of OCE in vascular diagnostics.

In conclusion, this study presents a systematic OCE-based approach for characterizing arterial stretch-dependent anisotropy, layer-specific inhomogeneity, and viscoelasticity. The results reveal a decreasing trend in arterial viscoelasticity and an increasing trend of layer-inhomogeneity with increasing mechanical stretch. The elastin network was found to exhibit obvious viscoelasticity at low strain. These findings provide new insights into cardiovascular biomechanics and could open the way for early-stage cardiovascular disease diagnosis and intervention strategies.

**METHODS**

*Sample preparation*

Fresh porcine descending thoracic aortas were obtained in a local abattoir and transported to lab on ice. Surrounding connective tissue was carefully removed. Square samples (2 × 2 cm) were cut with edges aligned along the circumferential and longitudinal directions of the arterial wall. Ten full-thickness aortic samples were prepared for measurements.

To investigate the specific contributions of collagen, five samples were further subjected to cyanogen bromide (CNBr) treatment, which effectively degrades collagen and other cellular and extracellular components while leaving elastin intact[17]. Briefly, samples were incubated in 50 mg/ml CNBr dissolved in 70% formic acid at room temperature for 19 h, followed by heating at 60 °C for 1 h and boiling for 5 min to deactivate the reagent. Treated samples were stored in 1x phosphate-buffered saline (PBS) until further experiments.

### *Optical coherence elastography (OCE)*

The OCE system was based on a swept-source OCT platform[46,47], utilizing a 1300 nm wavelength-swept laser with an 80 nm bandwidth operating at 43.2 kHz. The axial and lateral resolutions of the optical beam were approximately 15 μm and 30 μm, respectively. Galvanometer mirrors (Cambridge Technology, 6210H) enabled lateral scanning. A piezoelectric (PZT) actuator, coupled with a custom 3D-printed probe tip (2 mm wide, with an approximate 1 mm contact length), was used to generate surface vibrations in the sample. Pure-tone stimuli (1-20 kHz, $f_s$) excited harmonic waves in the tissue. For each OCE measurement, M-B scan mode was employed: 96 transverse positions (B-scan) were sampled, with 172 A-lines (M scan) recorded per position at 43.2 kHz. Fast Fourier Transform (FFT) analysis of M-scan profiles yielded the local amplitude and phase of displacements. Subsequent Fourier transformation in space revealed wave modes and wavenumbers. Phase velocity was calculated as $v = \omega/Re(k)$, where $\omega = 2\pi f_s$ and $k$ is the wavenumber.

Tissue samples were mounted on a custom biaxial stretcher. Carbon particles (~200 μm diameter) served as fiducial markers for stretch ratio calculations. OCE measurements were conducted along the axial and circumferential directions of the arterial samples at equibiaxial stretch ratios of 1 (stress-free), 1.1, 1.2, 1.3, and 1.4. OCE measurements were performed on the media side (air exposed), while the adventitial side was submerged in saline to prevent dehydration. Five measurements were performed at a single location for each stretch condition.

### *Analytic modeling of guided acoustic waves*

Acoustic waves guided along the arterial wall exhibit frequency-dependent dispersion characteristics[67,73]. When the tissue is under prestress, the dispersion relation is modulated via the acoustoelastic effect. To model this, we used the incremental dynamic theory of elasticity[53], where the wave equation is expressed as:

$$\nabla \cdot \mathbf{\Sigma} = \rho \mathbf{u}_{,tt}, \tag{7}$$

Here, $\boldsymbol{\Sigma}$ is the incremental stress tensor induced by acoustic waves with displacement $\boldsymbol{u}$, $\rho$ is the tissue mass density, and $t$ denotes time.

Given a strain energy function $W$ of deformation gradient tensor $\boldsymbol{F}$, the fourth-order Eulerian elasticity tensor is defined as $\mathcal{A}^0_{ijkl} = F_{iI} F_{kJ} \frac{\partial^2 W}{\partial F_{jI} \partial F_{lJ}}$ [53,74]. We adopted a Cartesian coordinate system with $x$ and $z$ axes in the tissue plane and the $y$ axis normal to the tissue surface. Under equibiaxial in-plane stretching ($\lambda_x = \lambda_z$), incompressibility yields $\lambda_y = (\lambda_x \lambda_z)^{-1}$, and $\boldsymbol{F} = \mathrm{diag}(\lambda_x, \lambda_y, \lambda_z)$.

To simplify the wave equation, we used a scalar stream function $\psi$, such that $u_x = \psi_{,y}$ and $u_y = -\psi_{,x}$. For harmonic guided waves propagating along the $x$-axis, the assumed form is:

$$\psi \propto e^{sky} e^{i(kx - \omega t)} \quad (8)$$

where $s$ is a complex decay parameter. Inserting Eq. (8) into the wave equation Eq. (7) yield a biquadratic equation in $s^2$, with two solutions $s_1^2$ and $s_2^2$. The full stream function is then written as $\psi = \sum_{i=-2}^{2} \psi_i e^{\mathrm{sign}(i) s_i k y} e^{i(kx - \omega t)}$.

In this study, we sequentially proposed single-layer elastic, single-layer viscoelastic, and two-layer viscoelastic models to progressively investigate the intrinsic viscoelastic tissue properties and layer-specific nature of the arterial wall. Boundary conditions are applied at the tissue-air and tissue-fluid interfaces. For the single-layer model, the upper surface is stress-free, and the lower surface maintains stress and displacement continuity with the fluid. For the two-layer model, additional continuity conditions are imposed at the media-adventitia interface. These boundary conditions lead to a secular equation:

$$\det(\mathbf{M}) = 0 \quad (9)$$

where $\mathbf{M}$ is a 5x5 matrix for single-layer and a 9x9 matrix for two-layer models. Explicit forms of $\mathbf{M}$ are detailed in Supplementary Notes 1 - 4. Solving this equation yields the phase velocities for the A0 and S0 Lamb wave modes. These correspond to the two lowest-order solutions without a frequency cutoff[75], exhibiting quasi-symmetric and quasi-antisymmetric displacement profiles due to asymmetric boundary conditions.

To extract mechanical parameters, dispersion curves were fitted using a least-squares error function: $\sqrt{\frac{1}{n} \sum_{i=1}^{n} \left( v_i^{(\mathrm{exp})} - v_i^{(\mathrm{model})} \right)^2}$, where $v_i^{(\mathrm{exp})}$ and $v_i^{(\mathrm{model})}$ are the experimental and model-predicted phase velocities, respectively. A genetic algorithm was employed for minimization. The parameter space used in the fitting was set to $1 \text{ kPa} < \alpha < 500 \text{ kPa}$, $0.1 < \gamma/\alpha \leq 1$, $1 \leq \beta/\alpha < 10$, $0 < \delta < 0.5$, and $0 < \eta < 0.05$ based on literature data[54,60,76].

### *Gasser-Ogden-Holzapfel (GOH) constitutive model*

To describe fiber-reinforced anisotropic elasticity in arteries, we employed the Gasser-Ogden-Holzapfel (GOH) model[77]. This model assumes two symmetrically distributed families of collagen fibers embedded in a non-fibrous matrix. The strain energy density function is:

$$W = \frac{\mu_0}{2}(I_1 - 3) + \frac{k_1}{k_2}\left(e^{k_2(\kappa I_1 + (1-3\kappa)I' - 1)^2} - 1\right) \tag{10}$$

Here, $\mu_0$ is the ground matrix shear modulus, $k_1$ is the fiber stiffness coefficient, and $k_2$ is a dimensionless exponent indicating nonlinear stiffening. The fiber dispersion parameter $\kappa$ ranges from 0 (aligned fibers) to 1/3 (random orientation). $I_1$ and $I'$ are the strain invariants defined as $I_1 = \lambda_c^2 + \lambda_r^2 + \lambda_a^2$ and $I' = \lambda_c^2 \cos^2\varphi + \lambda_a^2 \sin^2\varphi$. The invariants of the two fiber families, $I_4$ and $I_6$, are equal and thus combined as $I'$ (see details in Supplementary Note 5). $\lambda_r$, $\lambda_c$, and $\lambda_a$ are stretch ratios in the radial, circumferential, and axial directions, and $\varphi$ is the mean fiber angle.

The GOH model parameters were fitted to experimental values of $\alpha$, $\beta$ and $\gamma$, determined at multiple stretch ratios in both axial and circumferential directions (see Supplementary Note 5).

### *Finite element simulation*

Finite element simulation (Abaqus/CAE 6.14, Dassault Systèmes) was conducted to verify the mode mixture of A0 and S0 modes observed in the OCE experiments. The model included a thin plate (2 mm thickness) atop a fluid substrate with a plane-strain configuration. The plate's shear modulus was set to 100 kPa based on experimental data. Approximately 8000 plane-strain elements (CPE8RH) and 1600 acoustic elements (AC2D8) were used to discretize the tissue and fluid domains, respectively. Frequency analysis step was adopted to determine the modal shapes of the layer structure. Mesh convergence was verified via refinement tests. The spatial displacement fields under the A0 and S0 modes were extracted from the model and compared with the experimental results.

### *Statistical analysis*

All results are presented as mean ± standard deviation. Comparisons between groups were evaluated using unpaired Student's t-tests. A p-values less than 0.05 was considered statistically significant.

## Acknowledgements

This study was supported by funding from the National Institutes of Health via grants R01-HL098028, R01-EY027653, and R01-EY033356.

## Author contributions

Conceptualization: Y.Z. and S.-H.Y. Methodology: Y.J., G.-Y.L., R.W, and S.-H.Y. Investigation: Y.J., G.-Y.L., R.W., X.F. and S.-H.Y. Visualization: Y.J., G.-Y.L., and S.-H.Y. Funding acquisition: Y.Z. and S.-H.Y. Project administration: Y.Z. and S.-H.Y. Supervision: Y.Z. and S.-H.Y. Writing—original draft: Y.J., G.-Y.L., R.W., X.F., Y.Z. and S.-H.Y. Writing—review & editing: Y.J., Y.Z. and S.-H.Y.

## Data availability

Data underlying the results presented in this paper are not publicly available at this time but may be obtained from the authors upon reasonable request.

## Competing interests

The authors declare no competing interests.

## Supplementary materials

Supplementary materials associated with this article can be found in a document.

# Comprehensive characterization of nonlinear viscoelastic properties of arterial tissues using guided-wave optical coherence elastography

Yuxuan Jiang [1, †], Guo-Yang Li [1, 5, †], Ruizhi Wang [3, †], Xu Feng [1, 6, †],

Yanhang Zhang [3, 4, *], Seok-Hyun Yun [1, 2, *]

[1] Harvard Medical School and Wellman Center for Photomedicine, Massachusetts General Hospital, Boston, MA 02114, USA

[2] Harvard-MIT Health Sciences and Technology, Cambridge, MA 02139, USA

[3] Department of Mechanical Engineering, Boston University, Boston, MA 02215, USA

[4] Department of Biomedical Engineering, Boston University, Boston, MA 02215, USA

[5] Currently with the Department of Mechanics and Engineering Science, College of Engineering, Peking University, Beijing 100871, China

[6] Currently with the Department of Bioengineering, University of Texas at Dallas, TX 75080, USA

[†] Co-first authors with equal contribution.

[*] Corresponding authors: yanhang@bu.edu (Y.Z.), syun@hms.harvard.edu (S.H.Y.).

# Supplementary Figures

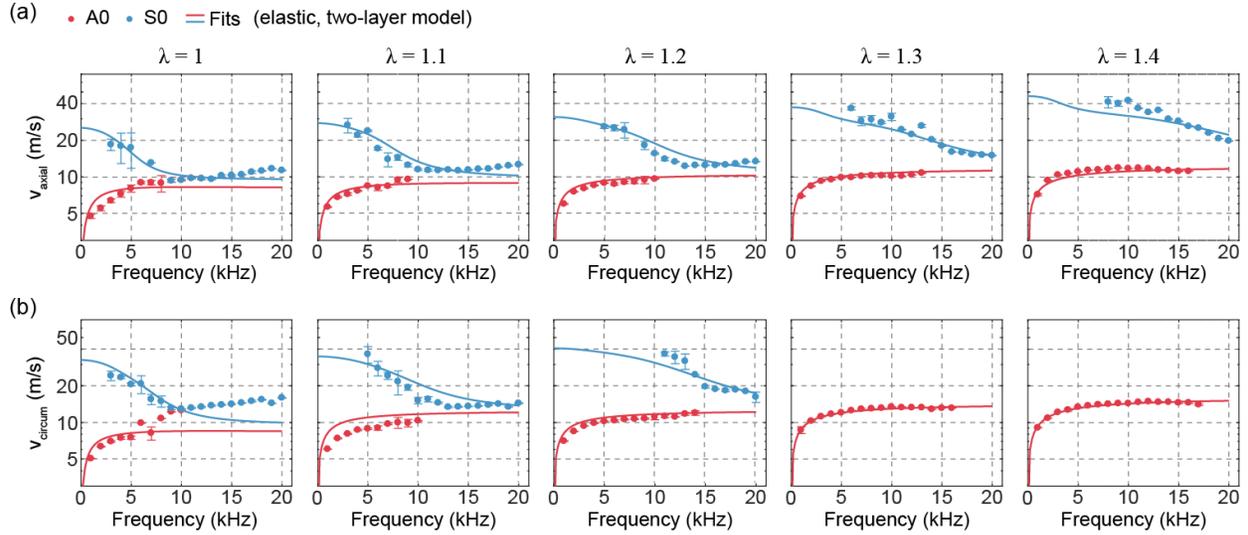

**Figure S1**. **Two-layer elastic model analysis of experimental data.** (a) Axial dispersion relations of A0 and S0 modes. (b) Circumferential dispersion relations of A0 and S0 modes. Markers: experiments. Lines: fitting curves using the bitwo-layer elastic model. The secular equation of the bitwo-layer elastic model is given by Eq. (S28) in Supplementary Note 4. The fitting parameters are listed in Supplementary Table S1.

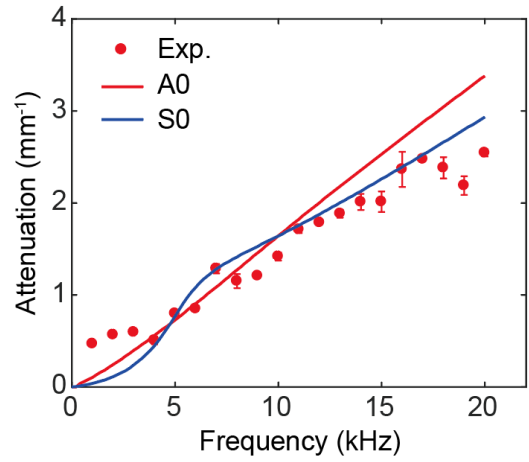

**Figure S2. Wave attenuation in the axial direction at λ = 1.** Markers: experiments. Lines: two-layer viscoelastic model-predicted wave attenuation for the A0 mode (red) and S0 mode (blue). Material parameters are provided in Table 3.

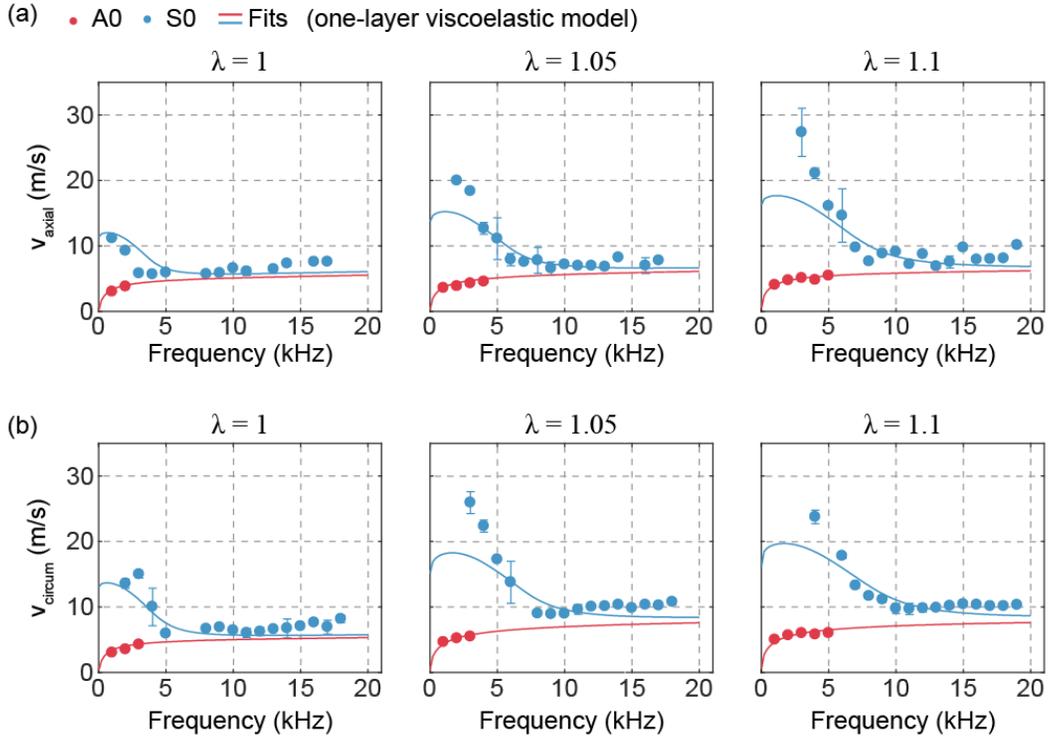

**Figure S3. Experimental dispersion of the artery after CNBr treatment.** (a) Axial data, (b) Circumferential data. Markers: experiments. Lines: fitting curves using the one-layer viscoelastic model.

## Supplementary Note 1. Derivation of the pre-stressed elastic single-layer model

Consider an elastic material that is subjected to finite deformation, and infinitesimal elastic waves are superimposed on the static deformation. The equation of wave motion is [1]

$$\nabla \cdot \mathbf{\Sigma} = \rho \mathbf{u}_{,tt}, \tag{S1}$$

where $\mathbf{\Sigma}$ denotes incremental stress induced by elastic waves. $\mathbf{u}$ denotes the displacement of wave motion. $\rho$ denotes the density of material. $t$ denotes the time. The subscript with a comma denotes partial differentiation with respect to the corresponding variable. The harmonic elastic wave can be described as $\mathbf{u} = \mathbf{u}_0 \exp(i(\mathbf{k} \cdot \mathbf{x} - \omega t))$. where $\mathbf{u}_0$, $\mathbf{k}$ and $\omega$ denote wave amplitude, wave vector, and angular frequency, respectively. For incompressible elastic materials, the incremental stress $\mathbf{\Sigma}$ is related to the displacement by

$$\Sigma_{ij} = \mathcal{A}^0_{ijkl} u_{l,k} - \hat{p}\delta_{ij} + p u_{i,j}, \quad i,j,k,l \in \{x,y,z\} \tag{S2}$$

where $\hat{p}$ denotes the increment of the Lagrange multiplier $p$. $\mathcal{A}^0_{ijkl}$ is the fourth-order Eulerian elasticity tensor defined as [1,2]

$$\mathcal{A}^0_{ijkl} = F_{iI} F_{kJ} \frac{\partial^2 W}{\partial F_{jI} \partial F_{lJ}}, \quad i,j,k,l,I,J \in \{x,y,z\} \tag{S3}$$

where $\mathbf{F} = \text{diag}(\lambda_x, \lambda_y, \lambda_z)$ is the deformation gradient tensor and $W$ is the strain energy function of the material. Inserting Eqs. (S2) and (S3) into Eq. (S1), the wave equation is [1,2]

$$-\hat{p}_{,j} + \mathcal{A}^0_{ijkl} u_{l,ik} = \rho u_{j,tt}, \quad i,j,k,l \in \{x,y,z\} \tag{S4}$$

Consider a flat plate with top side in contact with air and the bottom with fluid. The wall thickness of the plate is denoted as $h$. A Cartesian coordinate system (*x*, *y*, *z*) was established on the plate, where the *y*-axis denotes the thickness direction, and the *x*- and *z*-axes lies within the plane of the plate. The plate is subjected to in-plane biaxial stretch, with stretch ratios $\lambda_x$ and $\lambda_z$. Without loss of generality, we assume that waves in the plate propagate along the *x* direction, with displacement components confined to the *x-y* plane (i.e., $u_z = 0$), the stream function $\psi(x, y, t)$ can be used to replace displacements: $u_x = \psi_{,y}$ and $u_y = -\psi_{,x}$. Inserting $\psi$ into Eq. (S4), the wave equation becomes

$$\alpha \psi_{,xxxx} + 2\beta \psi_{,xxyy} + \gamma \psi_{,yyyy} = \rho(\psi_{,xxtt} + \psi_{,yytt}), \tag{S5}$$

where $\alpha$, $\beta$, and $\gamma$ are acoustoelastic parameters defined by $\alpha = \mathcal{A}^0_{xyxy}$, $2\beta = \mathcal{A}^0_{xxxx} + \mathcal{A}^0_{yyyy} - 2\mathcal{A}^0_{xxyy} - 2\mathcal{A}^0_{xyyx}$ and $\gamma = \mathcal{A}^0_{yxyx}$. Explicit forms of $\alpha$, $\beta$, and $\gamma$ are given in Supplementary Note 5. To solve Eq. (S5), a harmonic form of the stream function is assumed: $\psi = \psi_0 \exp(sky) \exp[i(kx - \omega t)]$, where $\psi_0$ is an amplitude; $s$ is a complex decay parameter; $\omega$ ($= 2\pi f$) is the angular frequency; $k$ is the complex wave number. Inserting the harmonic form of $\psi$ into Eq. (S5) yields

$$\gamma s^4 - \left(2\beta - \rho \frac{\omega^2}{k^2}\right) s^2 + \alpha - \rho \frac{\omega^2}{k^2} = 0. \tag{S6}$$

Four roots, denoted as $\pm s_1$ and $\pm s_2$, can be obtained from Eq. (S6). Thus, the general solution of $\psi$ takes the form: $\psi = \sum_{i=-2}^{2} \psi_i e^{sign(i) s_i k y} e^{i(kx - \omega t)}$.

The semi-infinite fluid at the bottom of the plate exhibits no rotational motion during the propagation of linear elastic waves, therefore a potential function $\phi$ is introduced to describe displacements of the fluid: $u_x^f = \phi_{,x}$, $u_y^f = \phi_{,y}$. The governing equation for acoustic waves in an inviscid stationary fluid is

$$\phi_{,xx} + \phi_{,yy} = \frac{1}{c_f^2} \phi_{,tt} \tag{S7}$$

where $c_f$ ($= \sqrt{\kappa_f / \rho_f}$) is the sound speed in the fluid. $\kappa_f$ denotes bulk modulus of the fluid. $\rho_f$ denotes the fluid density. The potential function also follows a harmonic form of $\phi = \phi_0 \exp(\xi k y) \exp[i(kx - \omega t)]$. $\xi$ can be solved by inserting $\phi$ into Eq. (S7), which yields

$$\xi^2 = 1 - \frac{1}{c_f^2} \frac{\omega^2}{k^2} \tag{S8}$$

The top surface of the plate exposed to air (at $y = h$) is stress free, and the bottom surface of the plate in contact with fluid (at $y = 0$) ensures continuity of normal displacement and stress. These boundary conditions can be written as [3]:

$$u_y = u_y^f, \Sigma_{yx} = 0, \Sigma_{yy} = -p^f, \text{at } y = 0;$$

$$\Sigma_{yx} = 0, \Sigma_{yy} = 0, \text{ at } y = h,$$

(S9)

where $u_i$ and $u_i^f$ denote the displacement of the plate and the fluid, respectively. $\Sigma_{ij}$ denotes the incremental stress of the plate. $p^f$ is the hydrostatic pressure of the fluid. With boundary conditions Eq. (S9), and replacing **u**, **u^f**, **Σ** and $p^f$ with $\psi$ and $\varphi$, we obtain the secular equation of the elastic single-layer model as follows

$$\det(\mathbf{M}_{5\times 5}^e) = 0 \qquad (S10)$$

where the components of the matrix $\mathbf{M}_{5\times 5}^e$ are

$M_{11} = M_{12} = M_{13} = M_{14} = 1, M_{15} = -i\xi,$

$M_{21} = 1 + s_1^2, M_{22} = 1 + s_1^2, M_{23} = 1 + s_2^2, M_{24} = 1 + s_2^2, M_{25} = 0,$

$M_{31} = \gamma s_1(1 + s_2^2), M_{32} = -\gamma s_1(1 + s_2^2), M_{33} = \gamma s_2(1 + s_1^2), M_{34} = -\gamma s_2(1 + s_1^2), M_{35} = i\rho_f \omega^2/k^2,$

$M_{41} = (1 + s_1^2)\exp(s_1 kh), M_{42} = (1 + s_1^2)\exp(-s_1 kh),$

$M_{43} = (1 + s_2^2)\exp(s_2 kh), M_{44} = (1 + s_2^2)\exp(-s_2 kh), M_{45} = 0,$

$M_{51} = s_1(1 + s_2^2)\exp(s_1 kh), M_{52} = -s_1(1 + s_2^2)\exp(-s_1 kh),$

$M_{53} = s_2(1 + s_1^2)\exp(s_2 kh), M_{54} = -s_2(1 + s_1^2)\exp(-s_2 kh), M_{55} = 0.$

(S11)

where $i$ in the element $M_{15}$ and $M_{35}$ denotes the imaginary unit.

## Supplementary Note 2. Derivation of the pre-stressed viscoelastic single-layer model

Consider a pre-stressed viscoelastic material subjected to linear elastic wave propagation; the wave equation still follows Eq. (S1). However, since the constitutive relation of the material has changed, the incremental stress for the incompressible viscoelastic material is given by [4]

$$\Sigma_{ij} = -\hat{q}\delta_{ij} + qu_{i,j} - G\hat{Q}\delta_{ij} + GQu_{i,j} + G\mathcal{A}^0_{ijkl}u_{l,k} - \Omega\sigma^e_{Dik}u_{j,k}, \quad (S12)$$
$$i,j,k,l \in \{x,y,z\}$$

where $\hat{q}$ denotes the increment of the Lagrange multiplier $q$. $\hat{Q}$ is the increment of the volumetric part of the elastic stress $Q$ ($= \sigma^e_{ii}/3$, summation with respect to $i$). The elastic Cauchy stress $\boldsymbol{\sigma}^e = (\partial W/\partial \boldsymbol{F})\boldsymbol{F}^{\mathrm{T}}$. $\mathcal{A}^0_{ijkl}$ is the fourth-order Eulerian elasticity tensor defined by Eq. (S3). $G$ and $\Omega$ in Eq. (S12) are two frequency-dependent parameters,

$$G = 1 + \eta(i\omega)^\delta, \quad \Omega = \eta(i\omega)^\delta. \quad (S13)$$

where $\eta$ (unit $s^\delta$) and $\delta$ (0<$\delta$<1, unit 1) are two viscoelastic parameters of the Kelvin-Voigt fractional derivative (KVFD) model [5,6]. $\eta$ denotes the ratio of material viscosity to elasticity. $\delta$ is a fractional order. When $\delta = 0$, it recovers to the elastic material; when $\delta = 1$, it recovers to the Kelvin-Voigt model (a spring and a dashpot in parallel). Inserting Eq. (S13) into Eq. (S1), the wave equation can be expressed by [4]

$$-\hat{q}_{,j} - G\hat{Q}_{,j} + G\mathcal{A}^0_{ijkl}u_{l,ik} - \Omega\sigma^e_{Dik}u_{j,ik} = \rho u_{j,tt}, \quad i,j,k,l \in \{x,y,z\} \quad (S14)$$

The viscoelastic single-layer model consists of a thin plate with a thickness of $h$, where its top surface in contact with air, and its bottom surface in contact with an inviscid fluid. Following the similar analysis in Supplementary Note 1, we introduce stream function $\psi$ to replace displacements— $u_x = \psi_{,y}$ and $u_y = -\psi_{,x}$. Inserting $\psi$ into Eq. (S14), we obtain the wave equation as follows

$$G(\alpha\psi_{,xxxx} + 2\beta\psi_{,xxyy} + \gamma\psi_{,yyyy}) - \Omega[\sigma_{Dxx}^e\psi_{,xxxx} + \sigma_{Dyy}^e\psi_{,yyyy} + (\sigma_{Dxx}^e + \sigma_{Dyy}^e)\psi_{,xxyy}] = \rho(\psi_{,xxtt} + \psi_{,yytt}),$$
(S15)

where the explicit forms of acoustoelastic parameters $\alpha$, $\beta$, and $\gamma$ are given in Supplementary Note 5. The deviatoric elastic stress $\sigma_{Dij}^e = \sigma_{ij}^e - Q$. The elastic stresses are related to the elasticity tensor as follows

$$Q = \frac{\alpha+\gamma+\mathcal{A}_{zyzy}^0}{3}, \quad \sigma_{Dxx}^e = \frac{2}{3}\alpha - \frac{\gamma}{3} - \frac{\mathcal{A}_{zyzy}^0}{3}, \quad \sigma_{Dyy}^e = \frac{2}{3}\gamma - \frac{\alpha}{3} - \frac{\mathcal{A}_{zyzy}^0}{3},$$
(S16)

The elastic waves are assumed to propagate along the x-axis, therefore the stream function follows the same harmonic form as adopted in the elastic model: $\psi = \psi_0\exp(sky)\exp[i(kx - \omega t)]$. Inserting the harmonic form of $\psi$ into Eq. (S15) yields

$$(G\gamma - \Omega\sigma_{Dyy}^e)s^4 + \left[\rho\frac{\omega^2}{k^2} - 2G\beta + \Omega(\sigma_{Dxx}^e + \sigma_{Dyy}^e)\right]s^2 + \left(G\alpha - \Omega\sigma_{Dxx}^e - \rho\frac{\omega^2}{k^2}\right) = 0.$$
(S17)

Four roots, denoted as $\pm s_1$ and $\pm s_2$, can be obtained from Eq. (S17). Thus, the general solution of $\psi$ takes the form: $\psi = \sum_{i=-2}^{2}\psi_i e^{sign(i)s_iky}e^{i(kx-\omega t)}$. With the boundary conditions Eq. (S9), and replacing **u**, **u**$^f$, **Σ** and $p^f$ with $\psi$ and $\varphi$, we obtain the secular equation of the viscoelastic single-layer model as follows

$$\det(\mathbf{M}_{5\times 5}^V) = 0$$
(S18)

where the components of the matrix $\mathbf{M}_{5\times 5}^V$ are

$M_{11} = M_{12} = M_{13} = M_{14} = 1, M_{15} = -i\xi,$

$M_{21} = 1 + s_1^2, M_{22} = 1 + s_1^2, M_{23} = 1 + s_2^2, M_{24} = 1 + s_2^2, M_{25} = 0,$

$M_{31} = C_1 s_1 - C_2 s_1^3 - \rho\frac{\omega^2}{k^2}s_1, \quad M_{32} = -(C_1 s_1 - C_2 s_1^3 - \rho\frac{\omega^2}{k^2}s_1),$

$$M_{33} = C_1 s_2 - C_2 s_2^3 - \rho \frac{\omega^2}{k^2} s_2, \ M_{34} = -(C_1 s_2 - C_2 s_2^3 - \rho \frac{\omega^2}{k^2} s_2), \ M_{35} = i\rho_f \frac{\omega^2}{k^2},$$

$$M_{41} = (1 + s_1^2)\exp(s_1 kh), M_{42} = (1 + s_1^2)\exp(-s_1 kh),$$

$$M_{43} = (1 + s_2^2)\exp(s_2 kh), M_{44} = (1 + s_2^2)\exp(-s_2 kh), M_{45} = 0,$$

$$M_{51} = \left(C_1 s_1 - C_2 s_1^3 - \rho \frac{\omega^2}{k^2} s_1\right)\exp(s_1 kh), \ M_{52} = -\left(C_1 s_1 - C_2 s_1^3 - \rho \frac{\omega^2}{k^2} s_1\right)\exp(-s_1 kh),$$

$$M_{53} = \left(C_1 s_2 - C_2 s_2^3 - \rho \frac{\omega^2}{k^2} s_2\right)\exp(s_2 kh), \ M_{54} = -\left(C_1 s_2 - C_2 s_2^3 - \rho \frac{\omega^2}{k^2} s_2\right)\exp(-s_2 kh), \ M_{55} = 0.$$

(S19)

where $C_1$ and $C_2$ are two coefficients defined as

$$C_1 = 2G\beta + \gamma + \Omega \mathcal{A}^0_{zyzy}, \ C_2 = \gamma + \frac{1}{3}\Omega(\alpha + \gamma + \mathcal{A}^0_{zyzy}), \tag{S20}$$

The symbol '$i$' in $M_{15}$ and $M_{35}$ denotes the imaginary unit. $\xi$ is defined by Eq. (S8). $\rho$ and $\rho_f$ denote the density of the plate and fluid, respectively. $\pm s_1$ and $\pm s_2$ are the roots of Eq. (S17).

One degenerate case can be validated: by inserting $G = 1$ and $\Omega = 0$ into Eq. (S18), the elastic single-layer model (i.e. Eq. (S10)) can be recovered.

## Supplementary Note 3. Derivation of the pre-stressed viscoelastic two-layer model

Now we consider a two-layer model, denoted as Layer 1 and Layer 2, with respective thickness $h_1$ and $h_2$. The top surface of Layer 1 is exposed to air, while the bottom surface of Layer 2 interfaces with a semi-infinite inviscid fluid. A Cartesian coordinate system (*x*, *y*, *z*) was established on the model, where the *y*-axis denotes the thickness direction, and the *x*- and *z*-axes denote the two directions parallel to the layers. Similar to the analysis in Supplementary Note 2, the stream function in Layer 1 follows a harmonic form of $\psi = \psi_0 \exp(sky)\exp[i(kx - \omega t)]$, and the stream function in Layer 2 follows a harmonic form of $\psi^* = \psi_0^* \exp(s^* ky)\exp[i(kx - \omega t)]$. Inserting $\psi$ into Eq. (S15), the wave equation in Layer 1 satisfies

$$(G_1\gamma_1 - \Omega_1 \sigma_{Dyy}^{e1})s^4 + \left[\rho_1 \frac{\omega^2}{k^2} - 2G_1\beta_1 + \Omega_1(\sigma_{Dxx}^{e1} + \sigma_{Dyy}^{e1})\right]s^2 +$$
$$\left(G_1\alpha_1 - \Omega_1 \sigma_{Dxx}^{e1} - \rho_1 \frac{\omega^2}{k^2}\right) = 0. \tag{S21}$$

Inserting $\psi^*$ into Eq. (S14), the wave equation in Layer 2 is

$$(G_2\gamma_2 - \Omega_2 \sigma_{Dyy}^{e2})s^{*4} + \left[\rho_2 \frac{\omega^2}{k^2} - 2G_2\beta_2 + \Omega_2(\sigma_{Dxx}^{e2} + \sigma_{Dyy}^{e2})\right]s^{*2} +$$
$$\left(G_2\alpha_2 - \Omega_2 \sigma_{Dxx}^{e2} - \rho_2 \frac{\omega^2}{k^2}\right) = 0. \tag{S22}$$

where $\rho_1$ and $\rho_2$ are material density of Layer 1 and Layer 2, respectively. $\sigma_{Dij}^{e1}$ and $\sigma_{Dij}^{e2}$ are stress of Layer 1 and Layer 2, respectively. $\alpha_1$, $\beta_1$, $\gamma_1$ are acoustoelastic parameters of Layer 1 (related to the elasticity tensor $\mathcal{A}_{ijkl}^0$ of Layer 1). $\alpha_2$, $\beta_2$, $\gamma_2$ are acoustoelastic parameters of Layer 2 (related to the elasticity tensor $\mathcal{A}_{ijkl}^{0*}$ of Layer 2). $G_1$ and $\Omega_1$ are frequency-dependent parameters of Layer 1. $G_2$ and $\Omega_2$ are frequency-dependent parameters of Layer 2. They are

$$G_1 = 1 + \eta_1(i\omega)^{\delta_1}, \quad \Omega_1 = \eta_1(i\omega)^{\delta_1},$$
$$G_2 = 1 + \eta_2(i\omega)^{\delta_2}, \quad \Omega_2 = \eta_2(i\omega)^{\delta_2}, \tag{S23}$$

where $\eta_1$ (relative strength of the viscosity compared to the elasticity) and $\delta_1$ (fractional order)

are the KVFD parameters of Layer 1. $\eta_2$ and $\delta_2$ are the KVFD parameters of Layer 2.

The interface of the two layers (at $y = 0$) ensures continuity of displacement and stress. The surface of Layer 1 exposed to air (at $y = h_1$) satisfies stress-free boundary conditions. The surface of Layer 2 exposed to the fluid (at $y = -h_2$) satisfies the continuity of the normal displacement and stress. These boundary conditions are expressed as

$$u_x = u_x^*, u_y = u_y^*, \Sigma_{yx} = \Sigma_{yx}^*, \Sigma_{yy} = \Sigma_{yy}^*, \quad \text{at } y = 0$$

$$\Sigma_{yx} = 0, \Sigma_{yy} = 0, \quad \text{at } y = h_1 \tag{S24}$$

$$u_y^* = u_y^f, \Sigma_{yx}^* = 0, \Sigma_{yy}^* = -p^f, \quad \text{at } y = -h_2$$

where $u_i$, $u_i^*$ and $u_i^f$ denote the displacement of Layer 1, Layer 2 and the fluid, respectively. $\Sigma_{ij}$ and $\Sigma_{ij}^*$ denote the incremental stress of Layer 1 and Layer 2, respectively. $p^f$ is the hydrostatic pressure of the fluid. Using $\psi$, $\psi^*$ and $\varphi$ to replace displacements and stresses in the boundary conditions Eq. (S24), we obtain the secular equation of the viscoelastic two-layer model as follows

$$\det(\mathbf{M}_{9\times 9}^V) = 0 \tag{S25}$$

where the components of the matrix $\mathbf{M}_{9\times 9}^V$ include

$M_{11} = (1 + s_1^2)\exp(s_1 k h_1), M_{12} = (1 + s_2^2)\exp(s_2 k h_1),$

$M_{13} = (1 + s_1^2)\exp(-s_1 k h_1), M_{14} = (1 + s_2^2)\exp(-s_2 k h_1), M_{15} = M_{16} = M_{17} = M_{18} = M_{19} = 0,$

$M_{21} = \left(C_1 s_1 - C_2 s_1^3 - \rho_1 \frac{\omega^2}{k^2} s_1\right)\exp(s_1 k h_1), M_{22} = \left(C_1 s_2 - C_2 s_2^3 - \rho_1 \frac{\omega^2}{k^2} s_2\right)\exp(s_2 k h_1),$

$M_{23} = -\left(C_1 s_1 - C_2 s_1^3 - \rho_1 \frac{\omega^2}{k^2} s_1\right)\exp(-s_1 k h_1), M_{24} = -\left(C_1 s_2 - C_2 s_2^3 - \rho_1 \frac{\omega^2}{k^2} s_2\right)\exp(-s_2 k h_1),$

$M_{25} = M_{26} = M_{27} = M_{28} = M_{29} = 0,$

$M_{31} = M_{32} = M_{33} = M_{34} = 0, M_{35} = \exp(-s_1^* k h_2), M_{36} = \exp(-s_2^* k h_2), M_{37} = \exp(s_1^* k h_2),$

$M_{38} = \exp(s_2^* k h_2), M_{39} = -i\xi \exp(-\xi k h_2),$

$M_{41} = M_{42} = M_{43} = M_{44} = 0, M_{45} = (1 + s_1^{*2})\exp(-s_1^* k h_2), M_{46} = (1 + s_2^{*2})\exp(-s_2^* k h_2),$

$M_{47} = (1 + s_1^{*2})\exp(s_1^* k h_2), \ M_{48} = (1 + s_2^{*2})\exp(s_2^* k h_2), \ M_{49} = 0,$

$M_{51} = M_{52} = M_{53} = M_{54} = 0, \ M_{55} = \left(C_1^* s_1^* - C_2^* s_1^{*3} - \rho_2 \frac{\omega^2}{k^2} s_1^*\right)\exp(-s_1^* k h_2),$

$M_{56} = \left(C_1^* s_2^* - C_2^* s_2^{*3} - \rho_2 \frac{\omega^2}{k^2} s_2^*\right)\exp(-s_2^* k h_2), \ M_{57} = -\left(C_1^* s_1^* - C_2^* s_1^{*3} - \rho_2 \frac{\omega^2}{k^2} s_1^*\right)\exp(s_1^* k h_2),$

$M_{58} = -\left(C_1^* s_2^* - C_2^* s_2^{*3} - \rho_2 \frac{\omega^2}{k^2} s_2^*\right)\exp(s_2^* k h_2), \ M_{59} = i\rho_f \frac{\omega^2}{k^2}\exp(-\xi k h_2),$

$M_{61} = s_1, M_{62} = s_2, M_{63} = -s_1, M_{64} = -s_2, M_{65} = -s_1^*, M_{66} = -s_2^*, M_{67} = s_1^*, M_{68} = s_2^*, M_{69} = 0,$

$M_{71} = M_{72} = M_{73} = M_{74} = 1, M_{75} = M_{76} = M_{77} = M_{78} = -1, M_{79} = 0,$

$M_{81} = C_2(1 + s_1^2), M_{82} = C_2(1 + s_2^2), M_{83} = C_2(1 + s_1^2), M_{84} = C_2(1 + s_2^2), M_{85} = -C_2^*(1 + s_1^{*2}),$

$M_{86} = -C_2^*(1 + s_2^{*2}), M_{87} = -C_2^*(1 + s_1^{*2}), M_{88} = -C_2^*(1 + s_2^{*2}), M_{89} = 0,$

$M_{91} = C_1 s_1 - C_2 s_1^3 - \rho_1 \frac{\omega^2}{k^2} s_1, \ M_{92} = C_1 s_2 - C_2 s_2^3 - \rho_1 \frac{\omega^2}{k^2} s_2, \ M_{93} = -(C_1 s_1 - C_2 s_1^3 - \rho_1 \frac{\omega^2}{k^2} s_1),$

$M_{94} = -(C_1 s_2 - C_2 s_2^3 - \rho_1 \frac{\omega^2}{k^2} s_2), \ M_{95} = -(C_1^* s_1^* - C_2^* s_1^{*3} - \rho_2 \frac{\omega^2}{k^2} s_1^*),$

$M_{96} = -(C_1^* s_2^* - C_2^* s_2^{*3} - \rho_2 \frac{\omega^2}{k^2} s_2^*), \ M_{97} = C_1^* s_1^* - C_2^* s_1^{*3} - \rho_2 \frac{\omega^2}{k^2} s_1^*,$

$M_{98} = C_1^* s_2^* - C_2^* s_2^{*3} - \rho_2 \frac{\omega^2}{k^2} s_2^*, \ M_{99} = 0.$ \hfill (S26)

where $i$ in the element $M_{39}$ and $M_{59}$ denotes the imaginary unit. $\xi$ is define by Eq. (S8). $\rho_1$, $\rho_2$, and $\rho_f$ denote the density of Layer 1, Layer 2, and the fluid, respectively. $\pm s_1$ and $\pm s_2$ are the roots solved by Eq. (S21). $\pm s_1^*$ and $\pm s_2^*$ are the roots solved by Eq. (S22). Coefficients $C_1$, $C_2$, $C_1^*$, and $C_2^*$ are defined by

$$C_1 = 2G_1\beta_1 + \gamma_1 + \Omega_1 \mathcal{A}_{zyzy}^0, \ C_2 = \gamma_1 + \frac{1}{3}\Omega_1(\alpha_1 + \gamma_1 + \mathcal{A}_{zyzy}^0),$$

$$C_1^* = 2G_2\beta_2 + \gamma_2 + \Omega_2 \mathcal{A}_{zyzy}^{0*}, \ C_2^* = \gamma_2 + \frac{1}{3}\Omega_2(\alpha_2 + \gamma_2 + \mathcal{A}_{zyzy}^{0*}).$$

(S27)

One degenerate case can be validated: by substituting $h = h_1 + h_2$ and assuming identical material properties for the two layers (i.e. $\alpha_1 = \alpha_2$, $\gamma_1 = \gamma_2$, $\beta_1 = \beta_2$, $\eta_1 = \eta_2$, $\delta_1 = \delta_2$, $\rho_1 = \rho_2$, etc.) into Eq. (S25), the viscoelastic single-layer model given in Eq. (S18) is recovered.

## Supplementary Note 4. Pre-stressed elastic two-layer model

The elastic two-layer guided wave model can be obtained as a special case of the viscoelastic two-layer model by inserting $G_1 = G_2 = 1$ and $\Omega_1 = \Omega_2 = 0$ into Eq. (S25), which yields

$$\det(\mathbf{M}^{\mathrm{e}}_{9\times 9}) = 0 \tag{S28}$$

where the nonzero components of the matrix $\mathbf{M}^{\mathrm{e}}_{9\times 9}$ include

$M_{11} = (1 + s_1^2)\exp(s_1 k h_1), M_{12} = (1 + s_2^2)\exp(s_2 k h_1), M_{13} = (1 + s_1^2)\exp(-s_1 k h_1),$

$M_{14} = (1 + s_2^2)\exp(-s_2 k h_1),$

$M_{21} = s_1(1 + s_2^2)\exp(s_1 k h_1), M_{22} = s_2(1 + s_1^2)\exp(s_2 k h_1),$

$M_{23} = -s_1(1 + s_2^2)\exp(-s_1 k h_1), M_{24} = -s_2(1 + s_1^2)\exp(-s_2 k h_1),$

$M_{35} = \exp(-s_1^* k h_2), M_{36} = \exp(-s_2^* k h_2), M_{37} = \exp(s_1^* k h_2), M_{38} = \exp(s_2^* k h_2),$

$M_{39} = -i\xi \exp(-\xi k h_2),$

$M_{45} = (1 + s_1^{*2})\exp(-s_1^* k h_2), M_{46} = (1 + s_2^{*2})\exp(-s_2^* k h_2),$

$M_{47} = (1 + s_1^{*2})\exp(s_1^* k h_2), M_{48} = (1 + s_2^{*2})\exp(s_2^* k h_2),$

$M_{55} = \gamma_2 s_1^*(1 + s_2^{*2})\exp(-s_1^* k h_2), M_{56} = \gamma_2 s_2^*(1 + s_1^{*2})\exp(-s_2^* k h_2),$

$M_{57} = -\gamma_2 s_1^*(1 + s_2^{*2})\exp(s_1^* k h_2), M_{58} = -\gamma_2 s_2^*(1 + s_1^{*2})\exp(s_2^* k h_2),$

$M_{59} = -i\rho_f \exp(-\xi k h_2)\omega^2/k^2,$

$M_{61} = s_1, M_{62} = s_2, M_{63} = -s_1, M_{64} = -s_2, M_{65} = -s_1^*, M_{66} = -s_2^*, M_{67} = s_1^*, M_{68} = s_2^*,$

$M_{71} = M_{72} = M_{73} = M_{74} = 1, M_{75} = M_{76} = M_{77} = M_{78} = -1,$

$M_{81} = \gamma_1(1 + s_1^2), M_{82} = \gamma_1(1 + s_2^2), M_{83} = \gamma_1(1 + s_1^2), M_{84} = \gamma_1(1 + s_2^2),$

$M_{85} = -\gamma_2(1 + s_1^{*2}), M_{86} = -\gamma_2(1 + s_2^{*2}), M_{87} = -\gamma_2(1 + s_1^{*2}), M_{88} = -\gamma_2(1 + s_2^{*2}),$

$M_{91} = \gamma_1 s_1(1 + s_2^2), M_{92} = \gamma_1 s_2(1 + s_1^2), M_{93} = -\gamma_1 s_1(1 + s_2^2), M_{94} = -\gamma_1 s_2(1 + s_1^2),$

$$M_{95} = -\gamma_2 s_1^*(1+s_2^{*2}), M_{96} = -\gamma_2 s_2^*(1+s_1^{*2}), M_{97} = \gamma_2 s_1^*(1+s_2^{*2}), M_{98} = \gamma_2 s_2^*(1+s_1^{*2}). \tag{S29}$$

where $\pm s_1$ and $\pm s_2$ are the four roots solved by the quartic equation

$$\gamma_1 s^4 - \left(2\beta_1 - \rho_1 \frac{\omega^2}{k^2}\right) s^2 + \alpha_1 - \rho_1 \frac{\omega^2}{k^2} = 0 \tag{S30}$$

$\pm s_1^*$ and $\pm s_2^*$ are the four roots solved by the quartic equation

$$\gamma_2 s^{*4} - \left(2\beta_2 - \rho_2 \frac{\omega^2}{k^2}\right) s^{*2} + \alpha_2 - \rho_2 \frac{\omega^2}{k^2} = 0 \tag{S31}$$

$\rho_1$, $\rho_2$ and $\rho_f$ are material density of the Layer 1, Layer 2 and fluid, respectively. $\alpha_1$, $\gamma_1$ and $\beta_1$ are acoustoelastic parameters of Layer 1. $\alpha_2$, $\gamma_2$ and $\beta_2$ are acoustoelastic parameters of Layer 2. $\xi$ is defined in Eq. (S8).

## Supplementary Note 5. Explicit forms of the acoustoelastic parameters

Gasser-Ogden-Holzapfel (GOH) constitutive model has been widely adopted to describe arterial hyperelasticity [7]. As shown in Fig. S4, we cut the tube longitudinally and unfold it into a flat plate. A Cartesian coordinate system $(x_r, x_c, x_a)$ is established on the plate, representing the radial (depth direction), circumferential, and axial direction of the artery, respectively (Fig. S4b). The strain energy function is

$$W = \frac{\mu_0}{2}(I_1 - 3) + \frac{k_1}{2k_2}\sum_{i=4,6}\{\exp[k_2(\kappa I_1 + (1-3\kappa)I_i - 1)^2] - 1\} \quad (S32)$$

The first term on the right-hand side of Eq. (S32) describes the isotropic elastin matrix. The second term describes anisotropic collagen fibers. $\mu_0$ denotes the matrix shear modulus. $k_1$ represents the collagen fiber-related shear modulus. $k_2$ is a dimensionless parameter denoting the nonlinear hardening effect of the collagen fibers. $\kappa$ is a fiber dispersion parameter, ranging from 0 for highly organized fibers to 1/3 random isotropic orientations. The first principle invariant $I_1 = \mathrm{tr}(F^T F)$, where $F$ is the deformation gradient tensor. $I_4$ and $I_6$ are two invariants related to two families of collagen fibers arranged along the preferred directions, $\boldsymbol{m_1}$ and $\boldsymbol{m_2}$, respectively. $I_4 = \boldsymbol{Fm_1} \cdot \boldsymbol{Fm_1}$ and $I_6 = \boldsymbol{Fm_2} \cdot \boldsymbol{Fm_2}$. This model assumes that the two families of collagen fibers are symmetrically distributed within the plane (Fig. S4), with orientations $\boldsymbol{m_1} = (0, \cos\varphi, \sin\varphi)^T$, and $\boldsymbol{m_2} = (0, -\cos\varphi, \sin\varphi)^T$, where $\varphi$ is the angle between the fiber orientations and the circumferential direction. Therefore $I_4 = I_6 = I'$, and the second term on the right-hand side of Eq. (S32) can be rewritten as a function of $I'$, resulting in the form shown in Eq. (10) in the main text.

Applying the above strain energy function into the definition of elasticity tensor $\mathcal{A}^0_{ijkl}$ (i.e. Eq. S3), we derive the explicit forms of the acoustoelastic parameters as follows:

$$\alpha_a = 2W_1\lambda_a^2 + 2W_4\lambda_a^2\sin^2\varphi + 2W_6\lambda_a^2\sin^2\varphi$$

$$\beta_a = W_1(\lambda_a^2 + \lambda_r^2) + W_4\lambda_a^2\sin^2\varphi + W_6\lambda_a^2\sin^2\varphi + 2W_{11}(\lambda_a^2 - \lambda_r^2)^2$$
$$+ 4W_{14}\lambda_a^2\sin^2\varphi(\lambda_a^2 - \lambda_r^2) + 4W_{16}\lambda_a^2\sin^2\varphi(\lambda_a^2 - \lambda_r^2)$$
$$+ 2W_{44}\lambda_a^4\sin^4\varphi + 2W_{66}\lambda_a^4\sin^4\varphi$$

$$\alpha_c = 2W_1\lambda_c^2 + 2W_4\lambda_c^2\cos^2\varphi + 2W_6\lambda_c^2\cos^2\varphi \tag{S33}$$

$$\beta_c = W_1(\lambda_c^2 + \lambda_r^2) + W_4\lambda_c^2\cos^2\varphi + W_6\lambda_c^2\cos^2\varphi + 2W_{11}(\lambda_c^2 - \lambda_r^2)^2$$
$$+ 4W_{14}\lambda_c^2\cos^2\varphi(\lambda_c^2 - \lambda_r^2) + 4W_{16}\lambda_c^2\cos^2\varphi(\lambda_c^2 - \lambda_r^2)$$
$$+ 2W_{44}\lambda_c^4\cos^4\varphi + 2W_{66}\lambda_c^4\cos^4\varphi$$

$$\gamma_a = \gamma_c = 2W_1\lambda_r^2$$

where $\alpha_a$ $(=\mathcal{A}^0_{arar})$, $\beta_a$ $(=(\mathcal{A}^0_{aaaa} + \mathcal{A}^0_{rrrr} - 2\mathcal{A}^0_{aarr} - 2\mathcal{A}^0_{arra})/2)$ and $\gamma_a$ $(=\mathcal{A}^0_{rara})$ correspond to the acoustoelastic parameters along the axial direction of the artery. $\alpha_c$ $(=\mathcal{A}^0_{crcr})$, $\beta_c$ $(=(\mathcal{A}^0_{cccc} + \mathcal{A}^0_{rrrr} - 2\mathcal{A}^0_{ccrr} - 2\mathcal{A}^0_{crrc})/2)$ and $\gamma_c$ $(=\mathcal{A}^0_{rcrc})$ are acoustoelastic parameters along the circumferential direction of the artery. $\lambda_r$, $\lambda_c$, $\lambda_a$ denote radial, circumferential, and axial stretch ratio, respectively. $W_i = \partial W / \partial I_i$, $W_{ij} = \partial^2 W / \partial I_i \partial I_j$, where $i,j = 1, 4, 6$. $I_1 = \lambda_c^2 + \lambda_r^2 + \lambda_a^2$. $I_4 = I_6 = \lambda_c^2\cos^2\varphi + \lambda_a^2\sin^2\varphi$. The parameter $\mathcal{A}^0_{zyzy}$ used in Eqs. (S20) and (S27) is equal to $\alpha_c$ when along the axial direction, while equal to $\alpha_a$ when along the circumferential direction.

In the stress-free state ($\lambda_c = \lambda_r = \lambda_a = 1$), Eq. (S33) reduces to $\alpha_a = \alpha_c = \mu_0$, $2\beta_a + 2\gamma_a = 4\mu_0 + 8k_1(1 - 3\kappa)^2\sin^2\varphi$, and $2\beta_c + 2\gamma_c = 4\mu_0 + 8k_1(1 - 3\kappa)^2\cos^2\varphi$. Therefore we have $(2\beta + 2\gamma)/\alpha > 4$ when $\kappa < 1/3$. Since the ratio $(2\beta + 2\gamma)/\alpha = 4$ corresponds to isotropy, and higher values indicate greater anisotropy [8], this result again suggests the anisotropic nature of arteries.

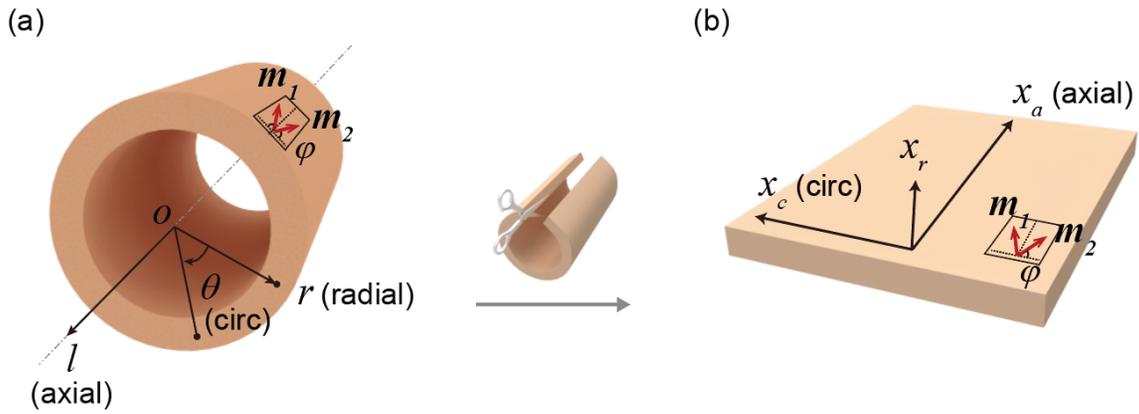

**Figure S4. Schematic of the Gasser-Ogden-Holzapfel constitutive model.** (a) A tube-shaped artery and (b) square arterial tissue sample with orientation axes and unit vectors of the two fiber families. $m_1$ and $m_2$ indicate orientation of the two symmetric fiber families, which induces the anisotropy of arteries.

## Supplementary Note 6. In-plane tensile modulus of a static plate

In the following, we consider a plate that is infinite in the $x_3$ direction, with $x_1$ as the longitudinal direction and $x_2$ as the thickness direction. After a finite pre-stretch, the plate further undergoes an incremental in-plane uniaxial tensile deformation along the $x_1$ direction. The following conditions are satisfied: displacement $u_3 = 0$, the normal stress $\Sigma_{22} = 0$, shear stress and shear strain components vanish, and all field variables are independent of the $x_3$ direction. With the above assumptions and the elastic stress-strain relation given in Eq. (S2), the incremental stresses are related to the displacements as follows:

$$\Sigma_{11} = \mathcal{A}^0_{1111}u_{1,1} + \mathcal{A}^0_{1122}u_{2,2} - \hat{p} + pu_{1,1} \tag{S34}$$

$$\Sigma_{22} = \mathcal{A}^0_{1122}u_{2,2} + \mathcal{A}^0_{2222}u_{2,2} - \hat{p} + pu_{2,2} \tag{S35}$$

The incompressible condition is $u_{1,1} + u_{2,2} = 0$. Using the stress assumption $\Sigma_{22} = 0$ into Eq. (S35), together with the incompressible relation, we can rewrite $\Sigma_{11}$ into the following form:

$$\Sigma_{11} = (\mathcal{A}^0_{1111} + \mathcal{A}^0_{2222} - 2\mathcal{A}^0_{1122} + 2p)u_{1,1} \tag{S36}$$

Since $\sigma_{22} = 0$ (Cauchy stress in the deformed state), the Lagrange multiplier $p$ is related to the acoustoelastic parameters as follows [2]

$$p = \mathcal{A}^0_{2121} - \mathcal{A}^0_{1221} \tag{S37}$$

Substituting Eq. (S37) into Eq. (S36), we can rewrite $\Sigma_{11}$ as

$$\Sigma_{11} = (\mathcal{A}^0_{1111} + \mathcal{A}^0_{2222} - 2\mathcal{A}^0_{1122} + 2\mathcal{A}^0_{2121} - 2\mathcal{A}^0_{1221})u_{1,1} \tag{S38-a}$$

The above equation is equivalent to

$$\Sigma_{11} = (2\beta + 2\gamma)u_{1,1} \tag{S38-b}$$

Eq. (S38-b) demonstrates the relationship between tensile stress and strain; therefore, $2\beta + 2\gamma$ corresponds to the in-plane tensile modulus.

## Supplementary Note 7. Complex dynamic modulus in pre-stressed viscoelastic materials

In viscoelastic materials, the shear wave velocity is determined by the complex shear modulus

$$\rho \frac{\omega^2}{k^2} = \mu^* \qquad (S39)$$

Inserting $s = 0$ into wave equation Eq. (S17), we can obtain $\mu^*$ as follows

$$\mu^* = G\alpha - \Omega \sigma^e_{Dxx} = G\alpha - \frac{1}{3}\Omega(2\alpha - \gamma - \mathcal{A}^0_{zyzy}) \qquad (S40)$$

The plate wave velocity is determined by the complex tensile modulus

$$\rho \frac{\omega^2}{k^2} = \bar{E}^* \qquad (S41)$$

In order to obtain the explicit form of $\bar{E}^*$ in pre-stressed viscoelastic material, we make use of incremental stress-strain relation given in Eq. (S12), and by following a similar derivation as shown in Supplementary Note 6, we get a relation $\Sigma_{xx} = \bar{E}^* u_{x,x}$ and

$$\bar{E}^* = G(2\beta + 2\gamma) - \Omega(\sigma^e_{Dxx} + \sigma^e_{Dyy} + 2q) = G(2\beta + 2\gamma) - \frac{1}{3}\Omega(-\alpha + 5\gamma - 4\mathcal{A}^0_{zyzy}) \qquad (S42)$$

Notably, unlike the linear case, the complex shear modulus is not simply $\mu^* = G\alpha$, and the complex tensile modulus is not directly $\bar{E}^* = G(2\beta + 2\gamma)$. This is primarily because, in the incremental dynamics framework, the material is first subjected to a fully relaxed finite pre-stress, and then viscoelastic wave motions are superimposed on top of it. As a result, some long-term Cauchy stress terms are retained in Eqs. (S40) and (S42).

Figure S5a compares the complex shear modulus $\mu^*$ and its first term $G\alpha$. The material parameters are based on the fitting results from experimental data. As shown, both the real and imaginary parts of $\mu^*$ are close to those of $G\alpha$ (relative error < 3% for the real part, and <14% for the imaginary part). Figure S5b compares the complex tensile modulus $\bar{E}^*$ and its first term $G(2\beta + 2\gamma)$. The relative difference between $\bar{E}^*$ and $G(2\beta + 2\gamma)$ is below 6% for the real part, and below 20% for the imaginary part. Therefore, in this study, it is reasonable to use $\mu^* \approx G\alpha$ and $\bar{E}^* \approx G(2\beta + 2\gamma)$ to approximate complex shear and tensile moduli, respectively, and we use these formulations to demonstrate how viscoelastic effects diminish with increasing pre-stress (as shown in Figs. 3 and 7).

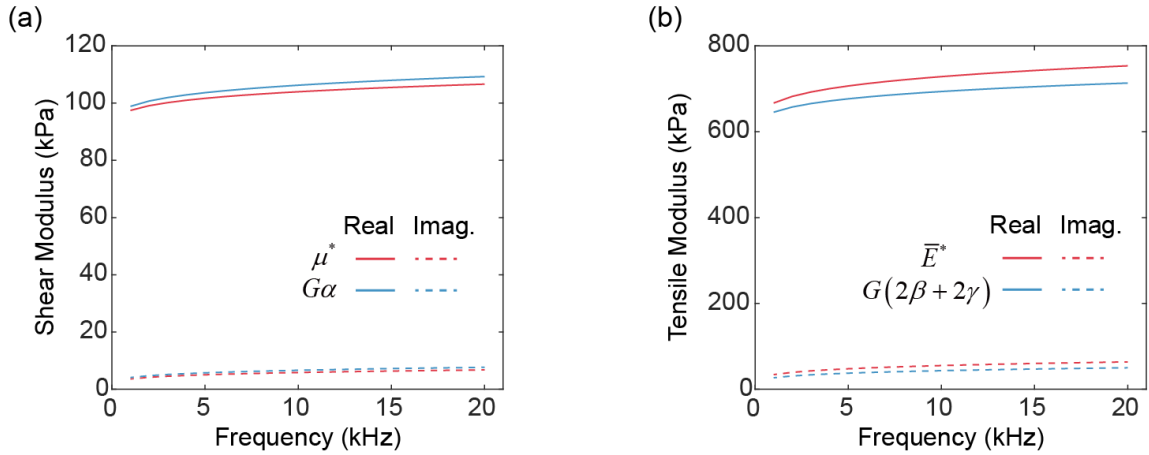

**Figure S5. Comparison of complex moduli and their approximate forms.** (a) Complex shear modulus $\mu^*$ and its first term $G\alpha$. (b) Complex tensile modulus $\bar{E}^*$ and its first term $G(2\beta + 2\gamma)$. The material parameters used here are obtained from the fitting results along the axial direction at $\lambda = 1.2$ (see Table 2), including $\alpha = 87$ kPa, $\gamma = 21$ kPa, $\beta = 263$ kPa, $\eta = 0.0023$, $\delta = 0.21$.

## Supplementary Note 8. Asymptotic phase velocities of the A0 and S0 modes at high frequencies in the two-layer model

### S8.1 Asymptotic solutions and their implication for experimental measurement

Figure S6 shows a schematic of the two-layer viscoelastic model, and representative dispersion curves of the A0 and S0 modes. We denote the high-frequency asymptotic phase velocities of the A0 and S0 modes as $c_{A0,\text{asymp}}$ and $c_{S0,\text{asymp}}$, respectively. They can be generally expressed by

$$c_{A0,\text{asymp}} = \min\{c_{R1}, c_{S2}\} \tag{S43}$$

$$c_{S0,\text{asymp}} = \text{second smallest of } \{c_{R1}, c_{t1}, c_{S2}, c_{t2}\} \tag{S44}$$

where $c_R$, $c_S$ and $c_t$ denote the Rayleigh surface wave, Scholte wave (fluid-solid interface wave), and plane shear wave velocity. The subscript $i$ (= 1, 2) denotes Layer $i$. For pre-stressed elastic materials, we denote $c_{R1} = n_{R1} c_{t1}$, and $c_{S2} = n_{S2} c_{t2}$, where $n_S$ and $n_R$ are dimensionless parameters less than 1. Based on the criterion given by Eqs. (S43) and (S44), the results can be divided into four regions according to the shear modulus ratio of the two layers ($\alpha_2/\alpha_1$), with each region corresponding to distinct asymptotic solutions:

- Region A: $\frac{\alpha_2}{\alpha_1} < n_{R1}^2$

$$c_{A0,\text{asymp}} = n_{S2} c_{t2}, \quad c_{S0,\text{asymp}} = c_{t2}$$

- Region B: $n_{R1}^2 \leq \frac{\alpha_2}{\alpha_1} < \left(\frac{n_{R1}}{n_{S2}}\right)^2$

$$c_{A0,\text{asymp}} = n_{S2} c_{t2}, \quad c_{S0,\text{asymp}} = n_{R1} c_{t1}$$

- Region C: $\left(\frac{n_{R1}}{n_{S2}}\right)^2 \leq \frac{\alpha_2}{\alpha_1} < \left(\frac{1}{n_{S2}}\right)^2$

$$c_{A0,\text{asymp}} = n_{R1} c_{t1}, \quad c_{S0,\text{asymp}} = n_{S2} c_{t2}$$

- Region D: $\frac{\alpha_2}{\alpha_1} \geq \left(\frac{1}{n_{S2}}\right)^2$

$$c_{A0,\text{asymp}} = n_{R1} c_{t1}, \quad c_{S0,\text{asymp}} = c_{t1}$$

For linear elastic and isotropic materials, $n_S = 0.839$ and $n_R = 0.955$, and the ranges of the four regions reduce to $\mu_2/\mu_1 \in (0, 0.912)$, $[0.912, 1.296)$, $[1.296, 1.421)$, and $[1.421, +\infty)$, respectively. For pre-stressed elastic materials, $n_R$ and $n_S$ can be determined by solving the following two equations. The Rayleigh wave equation in the pre-stressed elastic material is [9]

$$s_1(1 + s_2^2)^2 - s_2(1 + s_1^2)^2 = 0 \tag{S45}$$

and the Scholte wave equation is [10]

$$\gamma s_2(1 + s_1^2)^2 - \gamma s_1(1 + s_2^2)^2 + (s_2^2 - s_1^2)\frac{\rho_f}{\xi}\frac{\omega^2}{k^2} = 0 \tag{S46}$$

where $\xi$ is defined in Eq. (S8). $s_2$ and $s_2$ are roots solved by Eq. (S6). $\rho_f$ is the fluid density. Eqs. (S45) and (S46) indicate that $n_S$ and $n_R$ are functions of the acoustoelastic parameters $\alpha$, $\gamma$, and $\beta$. It can be shown that $n_R$ increase with the stretch ratio and approaches 1. The variation of $n_S$ is model-dependent; for the GOH model, it also increases with increasing stretch.

By fitting to the experimental dispersion, we have obtained the acoustoelastic parameters $\alpha_i$, $\gamma_i$, and $\beta_i$ ($i = 1, 2$) at multiple stretch ratios ($\lambda = 1 \sim 1.4$). With the help of Eqs. (S45) – (S46), the values of $n_{S2}$ and $n_{R1}$ at each stretching state can be solved, and then the boundaries of the four regions (i.e. values of $n_{R1}^2$, $(n_{R1}/n_{S2})^2$, and $1/n_{R1}^2$) at the corresponding stretching state can be determined. The results are shown in Fig. S7. For both axial and circumferential directions, when $\lambda = 1$, the shear modulus ratio of the two layers ($\alpha_2/\alpha_1$) falls within Region A. When $\lambda \geq 1.1$, the ratios shift into Region D. Therefore, in the stress-free state, the asymptotic solutions of A0 and S0 modes are governed by the adventitia, with $c_{A0,\text{asymp}} = n_{S2}c_{t2}$ and $c_{S0,\text{asymp}} = c_{t2}$, whereas after stretching, the asymptotic solutions are governed by the media, with $c_{A0,\text{asymp}} = n_{R1}c_{t1}$, and $c_{S0,\text{asymp}} = c_{t1}$.

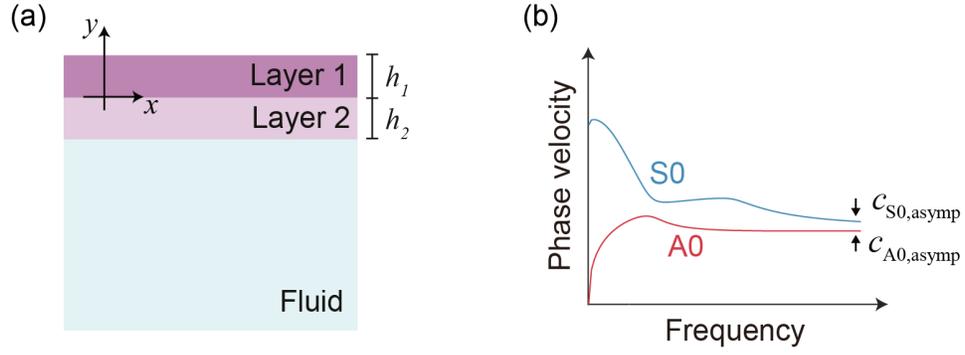

**Figure S6. Two-layer guided wave model.** (a) Schematic of the model. (b) A0 and S0 modes of the two-layer viscoelastic model. $c_{A0,\text{asymp}}$ and $c_{S0,\text{asymp}}$ denote the high-frequency asymptotic solutions for the two modes.

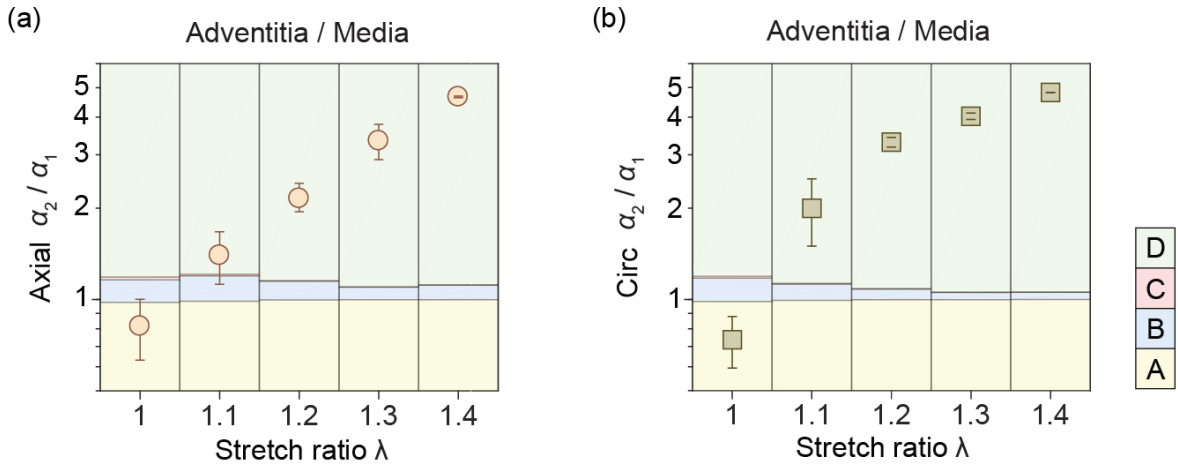

**Figure S7. Plot of shear modulus ratio ($\alpha_2/\alpha_1$) versus stretch ratios, and classification of the four regions at the corresponding stretch ratio.** (a) Axial data, (b) Circumferential data.

### S8.2 Critical frequencies for the asymptotic velocities of the A0 and S0 modes

The critical frequencies of the asymptotic velocities for the A0 and S0 modes are denoted as $f_{c,A0}$, and $f_{c,S0}$, and they can be approximately estimated as follows:

- Region A: $f_{c,A0} = \frac{c_{S2}}{h_2}$, $f_{c,S0} = \frac{2.5 c_{t2}}{h_2}$

- Region B: $f_{c,A0} = \frac{2 c_{S2}}{h_1+h_2}$, $f_{c,S0} = \frac{2 c_{R1}}{h_1+h_2}$

- Region C: $f_{c,A0} = \frac{c_{R1}}{h_1}$, $f_{c,S0} = \frac{2.5 c_{S2}}{h_2}$

- Region D: $f_{c,A0} = \frac{c_{R1}}{h_1}$, $f_{c,S0} = \frac{2.5 c_{t1}}{h_1}$

Taking the axial data for $\lambda = 1.4$, for example, the shear modulus of the media is $\alpha_1 = 66$ kPa, and $c_{t1} = 8.2$ m/s. The wall thickness is $h_1 = 0.3$ mm. Using the critical frequencies defined in Region D, we get $f_{c,A0} = 25$ kHz, and $f_{c,S0} = 65$ kHz. Especially, $f_{c,S0}$ is much higher than 20 kHz, indicating that the current OCE only captures the intermediate frequency range of S0 mode. Higher-frequency measurements may reveal more mechanical information of artery samples.

**Supplementary Note 9. Literature data of shear and tensile moduli of artery tissues**

We compare our characterization results of the shear and tensile moduli of arterial samples with those reported in the literature. The shear and tensile moduli under different stretch ratios were calculated based on the constitutive parameters of arteries reported in the literature. Figure S8a presents predictions derived from the data of Giudici et al [11], and Fig. S8b shows predictions based on the data of Sommer et al [12]. Figure S8-a1 and b1 shows the bidirectional shear moduli of the media and the adventitia. Figure S8-a2 and b2 shows the bidirectional tensile moduli of the media and the adventitia. In general, these moduli increase with respect to the stretch ratio. The circumferential moduli are higher than the axial ones. Figure S8-a3 and b3 plot the ratio of shear modulus of the two layers. Figure S8-a4 and b4 plot the ratio of tensile modulus of the two layers. These modulus ratios are lower than 1 when the artery is stress-free, indicating a stiffer media in this state. With the increase of the stretch ratio, these ratios increase gradually and surpass 1, indicating that the adventitia becomes significantly stiffer under stretching. These result align well with our experimental results.

The moduli or modulus ratios predicted in the literature at higher stretch ratios (e.g., >1.2) are significantly larger than the values obtained from our experimental fitting. In fact, the stiffening coefficient of collagen fibers reported in the literature is typically above 10, whereas our fitting results yield a fiber stiffening coefficient ($k_2$) of only around 4. One possible explanation is that the stretch ratio measured in our experiments represents the average value over a broader sample area compared to the area of wave propagation. The localized stretch ratio in the regions where elastic waves propagate could be smaller. The overestimation of the stretch ratio results in an underestimation of the fiber stiffening coefficient.

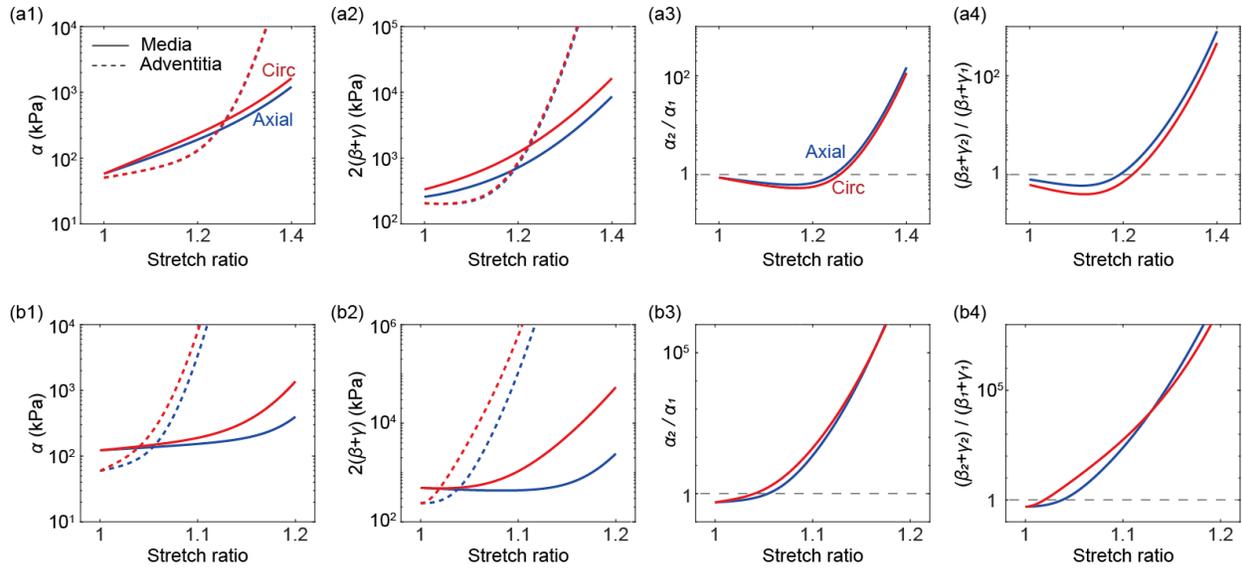

**Figure S8. Literature data of shear and tensile moduli of the artery tissues.** (a1) - (a4): constitutive parameters obtained from Giudici et al [11]. Tested tissue: porcine thoracic aortas. (b1) – (b4): constitutive parameters obtained from Sommer et al [12]. Tested tissue: human carotid arteries. (a1) and (b1), Bidirectional and bilayers' shear moduli with respect to the stretch ratio. (a2) and (b2), Bidirectional and bilayers' tensile moduli with respect to the stretch ratio. (a3) and (b3), Bidirectional ratios of the shear moduli of the two layers. (a4) and (b4), Bidirectional ratios of the tensile moduli of the two layers.

**Supplementary Tables**

Table S1. Measured modulus parameters from the two-layer elastic model (1-20 kHz)

|  |  | $\lambda = 1.0$ | $\lambda = 1.1$ | $\lambda = 1.2$ | $\lambda = 1.3$ | $\lambda = 1.4$ |
|---|---|---|---|---|---|---|
| Axial, intima-media | $\alpha$ (kPa) | 94 ± 4 | 97 ± 5 | 102 ± 10 | 116 ± 9 | 118 ± 8 |
|  | $2\beta + 2\gamma$ (kPa) | 600 ± 180 | 630 ± 60 | 680 ± 150 | 750 ± 100 | 1050 ± 140 |
| Axial, adventitia | $\alpha$ (kPa) | 82 ± 10 | 89 ± 9 | 130 ± 20 | 170 ± 80 | 190 ± 35 |
|  | $2\beta + 2\gamma$ (kPa) | 580 ± 140 | 890 ± 120 | 1250 ± 300 | 2050 ± 350 | 3200 ± 450 |
| Circum., intima-media | $\alpha$ (kPa) | 96 ± 1 | 140 ± 12 | 145 ± 8 | 165 ± 7 | 190 ± 1 |
|  | $2\beta + 2\gamma$ (kPa) | 1030 ± 100 | 1200 ± 100 | 1380 ± 120 | - | - |
| Circum., adventitia | $\alpha$ (kPa) | 80 ± 10 | 170 ± 40 | 190 ± 60 | 260 ± 60 | 350 ± 20 |
|  | $2\beta + 2\gamma$ (kPa) | 870 ± 80 | 1200 ± 520 | 1950 ± 220 | - | - |